\documentclass[twocolumn]{aastex62}
\usepackage{natbib}
\graphicspath{{./}{figures/}}
\usepackage{color,subfigure,lineno,booktabs} 

\definecolor{firebrick}{rgb}{0.7, 0.13, 0.13}

\newcommand{\Hb}{H$\beta$}
\newcommand{\Ha}{H$\alpha$}
\newcommand{\Pab}{Pa$\beta$}

\newcommand{\zphot}{$z_{\rm phot}$}
\def\CIV{C\,{\sc iv}}
\def\CIII{C\,{\sc iii]}}

\def\HeI{He\,{\sc i}}
\def\CIV{C\,{\sc iv}}
\def\MgII{Mg\,{\sc ii}}

\def\OIII{[O\,{\sc iii}]}
\def\NII{[N\,{\sc ii}]}

\def\OII{[O\,{\sc ii}]}
\def\NeIII{[Ne\,{\sc iii}]}
\def\SIII{[S\,{\sc iii}]}

\def\proj{\texttt{NEXUS}}

\shorttitle{NEXUS: Quick Release Notes} %
\shortauthors{Zhuang et~al.}

\begin{document}

\title{NEXUS: Notes on Quick Data Releases}


\author[0000-0001-5105-2837]{Ming-Yang Zhuang}
\affiliation{Department of Astronomy, University of Illinois Urbana-Champaign, Urbana, IL 61801, USA}
\email{mingyang@illinois.edu}

\author[0000-0003-1659-7035]{Yue Shen}
\affiliation{Department of Astronomy, University of Illinois Urbana-Champaign, Urbana, IL 61801, USA}
\affiliation{National Center for Supercomputing Applications, University of Illinois Urbana-Champaign, Urbana, IL 61801, USA}

\author[0000-0002-1605-915X]{Junyao Li}
\affiliation{Department of Astronomy, University of Illinois Urbana-Champaign, Urbana, IL 61801, USA}

\author[0000-0003-0230-6436]{Zhiwei Pan}
\affiliation{Department of Astronomy, University of Illinois Urbana-Champaign, Urbana, IL 61801, USA}

\author[0000-0001-7201-1938]{Lei Hu}
\affiliation{Department of Physics and Astronomy, University of Pennsylvania, Philadelphia, 209 South 33rd Street, PA 19104, USA}

\author[0000-0002-6523-9536]{Adam J.\ Burgasser}
\affiliation{Department of Astronomy \& Astrophysics, UC San Diego, La Jolla, CA, USA}

\author[0000-0003-4263-2228]{David~A.~Coulter}
\affiliation{William H. Miller III Department of Physics and Astronomy, Johns Hopkins University, 3400 North Charles Street, Baltimore, MD 21218, USA.}
\affiliation{Space Telescope Science Institute, 3700 San Martin Drive, Baltimore, MD 21218, USA.}

\author[0000-0002-5612-3427]{Jenny E. Greene}
\affiliation{Department of Astrophysical Sciences, Princeton University, 4 Ivy Lane, Princeton, NJ 08544, USA}

\author[0000-0002-7633-431X]{Feige Wang}
\affiliation{Steward Observatory, University of Arizona, Tucson, AZ 85750, USA}

\author{the NEXUS team}

\begin{abstract}
NEXUS is a JWST Multi-Cycle (Cycles 3--5) GO Treasury imaging and spectroscopic survey around the North Ecliptic Pole during 2024--2028. It contains two overlapping tiers in depth and area coverage. The Wide tier ($\sim 400~{\rm arcmin}^2$) performs NIRCam/WFSS 2.4--5\,\micron\ grism spectroscopy with three annual epochs over 3 years (final spectral continuum ${\rm S/N/pixel>3}$ at F444W$<22.2$), accompanied by NIRCam multi-band imaging in F090W, F115W, F150W, F200W, F356W and F444W. The Deep tier ($\sim 50~{\rm arcmin}^2$) performs high-multiplexing NIRSpec {0.54--5.5}\,\micron\ MOS/PRISM spectroscopy for $\sim 10,000$ targets in total, over 18 epochs with a 2-month cadence, along with F200W+F444W NIRCam imaging for each epoch. Parallel imaging observations with MIRI and additional NIRCam filters are also performed within the Wide and Deep tiers. The primary data covering the Deep tier (including NIRCam imaging, NIRSpec/MSA spectra, and vetted MSA spectroscopic redshifts) are released in regular Quick Data Releases to facilitate follow-up studies. This evolving document describes the MSA targeting information and observing status for each of the 18 Deep epochs, which started in May 2025 and continue on the regular 2-month cadence. We also describe the content and caveats of the quick release data and report selected cases of diverse scientific interests.  
\end{abstract}

\section{Introduction}\label{sec:introduction}


\proj\ (PID: 5105) is a multi-cycle (Cycles 3-5) JWST imaging and spectroscopic survey around the North Ecliptic Pole. It features two overlapping tiers. The \textbf{Wide tier} covers a footprint of $\sim 400\,{\rm arcmin^2}$, and performs NIRCam imaging in six filters (F090W, F115W, F150W, F200W, F356W, F444W) and WFSS slitless spectroscopy in F322W2 and F444W. The Wide tier is revisited annually over three cycles with a $\Delta t$ of about 17 months (PA orientation about $\pm 150$~deg). However, due to additional scheduling constraints, the first Wide epoch was split in two, with the central $\sim 100\,{\rm arcmin^2}$ observed in September 2024 \citep{nexus_edr} and the rest of Wide Epoch 1 observed in June 2025. The \textbf{Deep tier} covers the central $\sim 50\ {\rm arcmin^2}$ within the Wide tier, and performs NIRSpec MSA/PRISM spectroscopy and F200W+F444W NIRCam imaging on a 2-month cadence through early 2028. For both Wide and Deep observations, there are coordinated parallel observations with MIRI imaging and NIRCam imaging in additional filters, but these parallel observations have smaller coverage than the primary observations. The schedule for the multi-epoch Wide and Deep observations is now fixed, as shown in {Table~\ref{tab:schedule}}. 

The full details of the program design and science cases are presented in the NEXUS overview paper \citep{nexus}. NEXUS will produce annual data releases on the data taken in the previous cycle. However, given the nature of the cadenced Deep observations, it is of great value to present quick data releases on the Deep observations to guide prompt science analyses and follow-up observations. This document and the associated data products serve the latter purpose. We describe the top-level information about the Deep epoch observations and quick data releases in Section~\ref{sec:deep}. In Section~\ref{sec:deep_notes} we provide details for each individual Deep epochs. This document will be regularly updated throughout the NEXUS survey to include the most recent Deep epoch, typically $\sim 2$~months after the new data are taken. 

\begin{deluxetable}{ccccc}
\tablecaption{NEXUS Observation Schedule \label{tab:schedule}}
\tablewidth{0pt}
\tablehead{
\colhead{Observation} & \colhead{Plan Window/Execution Date}
}
\startdata
Deep-epoch1  &  2025/6/1 \\
Deep-epoch2  &  2025/7/30 \\
Deep-epoch3  &  2025/9/28 \\
Deep-epoch4  &  2025/11/28 \\
Deep-epoch5  &  2026/1/28 \\
Deep-epoch6  & 2026/3/28   \\
Deep-epoch7  & 2026/5/28   \\
Deep-epoch8  & 2026/8/7 \\
Deep-epoch9  & 2026/9/28 \\
Deep-epoch10 & 2026/11/21 -- 2026/11/28 \\
Deep-epoch11 & 2027/1/20  -- 2027/1/27  \\
Deep-epoch12 & 2027/3/21  -- 2027/3/28 \\
Deep-epoch13 & 2027/5/20  -- 2027/5/27 \\
Deep-epoch14 & 2027/7/19  -- 2027/7/26  \\
Deep-epoch15 & 2027/9/17  -- 2027/9/24  \\
Deep-epoch16 & 2027/11/16 -- 2027/11/23 \\
Deep-epoch17 & 2028/1/15  -- 2028/1/22  \\
Deep-epoch18 & 2028/3/15  -- 2028/3/22 \\
\hline
Wide-epoch1-center  &  2024/9/12       \\
Wide-epoch1-outer  & 2025/6/23       \\
Wide-epoch2  & 2026/2/5  -- 2026/2/12        \\
Wide-epoch3  & 2027/7/3   -- 2027/7/11         \\
\enddata
\end{deluxetable}


\section{Overview of Deep observations}\label{sec:deep}

\subsection{Overview of MSA targeting scheme}\label{sec:msa_gen}

NEXUS-Deep targets photometric objects for NIRSpec MSA spectroscopy with PRISM. Each Deep epoch has four NIRSpec/MSA pointings to cover 0.6--5.3 \micron. We adopt the \texttt{NRSIRS2RAPID} readout pattern, 43 Groups, 1 Integration, and a two-shutter slitlet nodding for each pointing, with an effective exposure time of 21.4 min for each target. 

Targets of different science categories are selected from the photometric catalog, compiled using all available NEXUS and ground-based photometry. We classify the objects into seven classes based on science category, each accompanied by a set of criteria on magnitude, photometric redshift (\zphot), morphology, and colors. Each class has its own weight (higher weights mean higher priority in allocating an MSA slit), allowing us to control their observing probability and priority. The descriptions of each class are the following:

\begin{figure*}[t]
    \centering
    \includegraphics[width=0.33\linewidth]{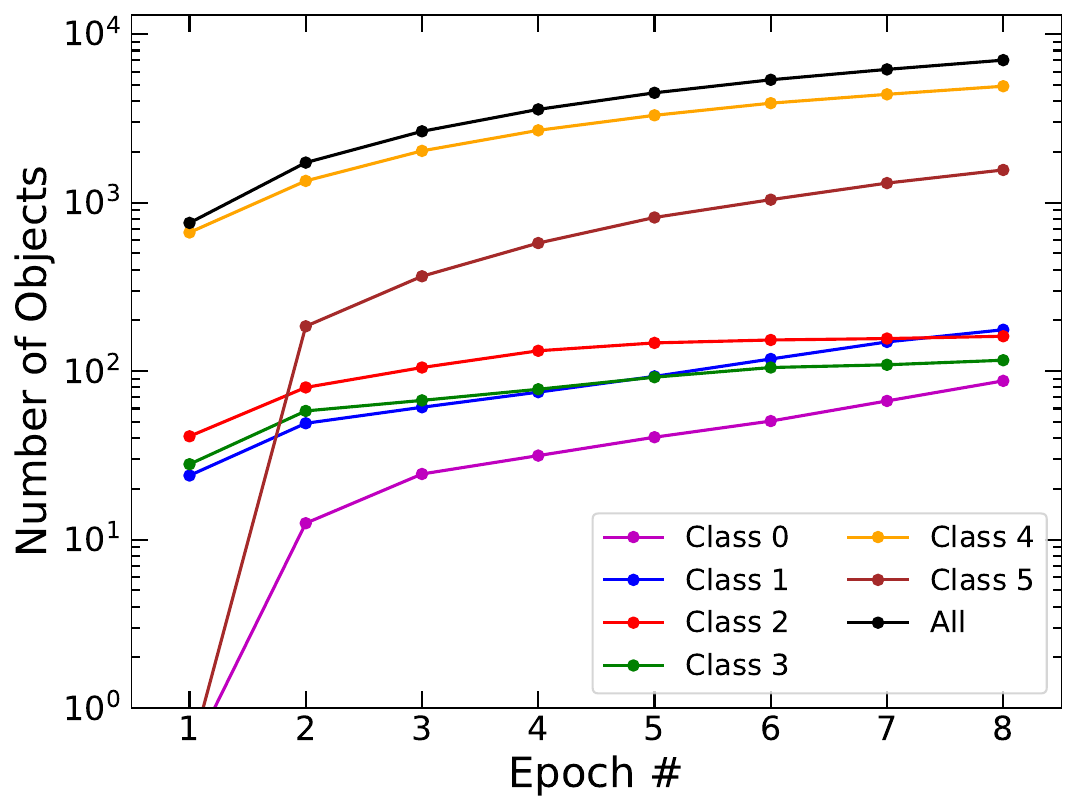}
    \includegraphics[width=0.31\linewidth]{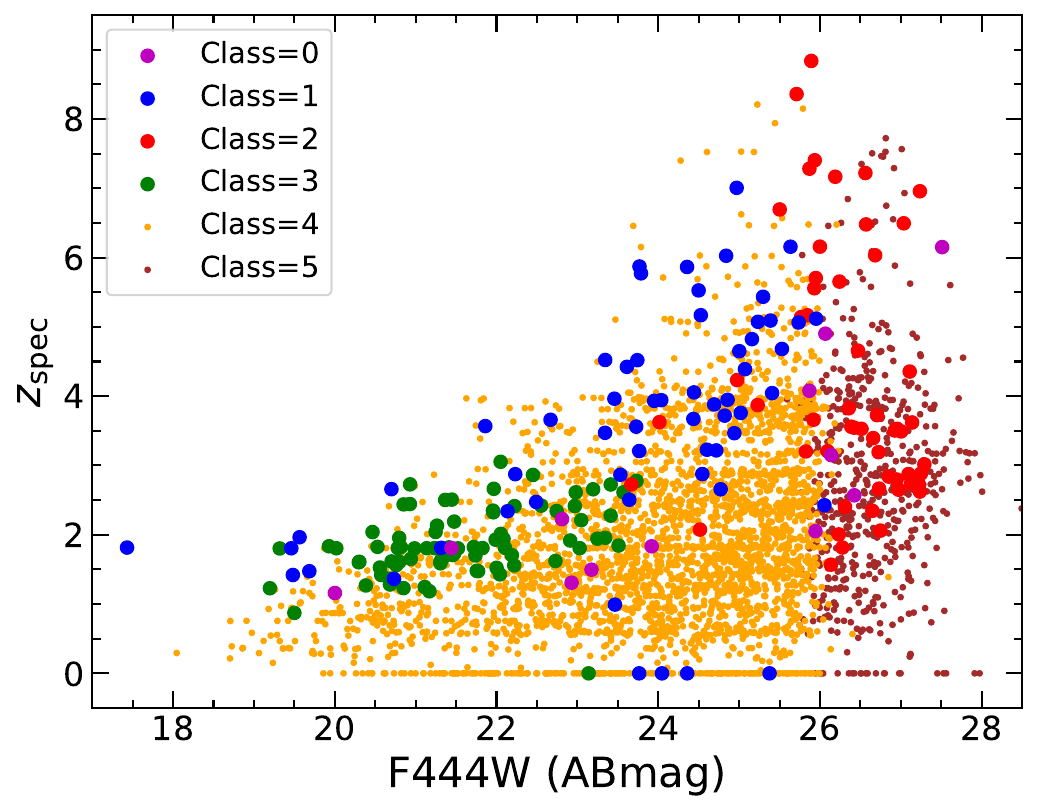}
    \includegraphics[width=0.33\linewidth]{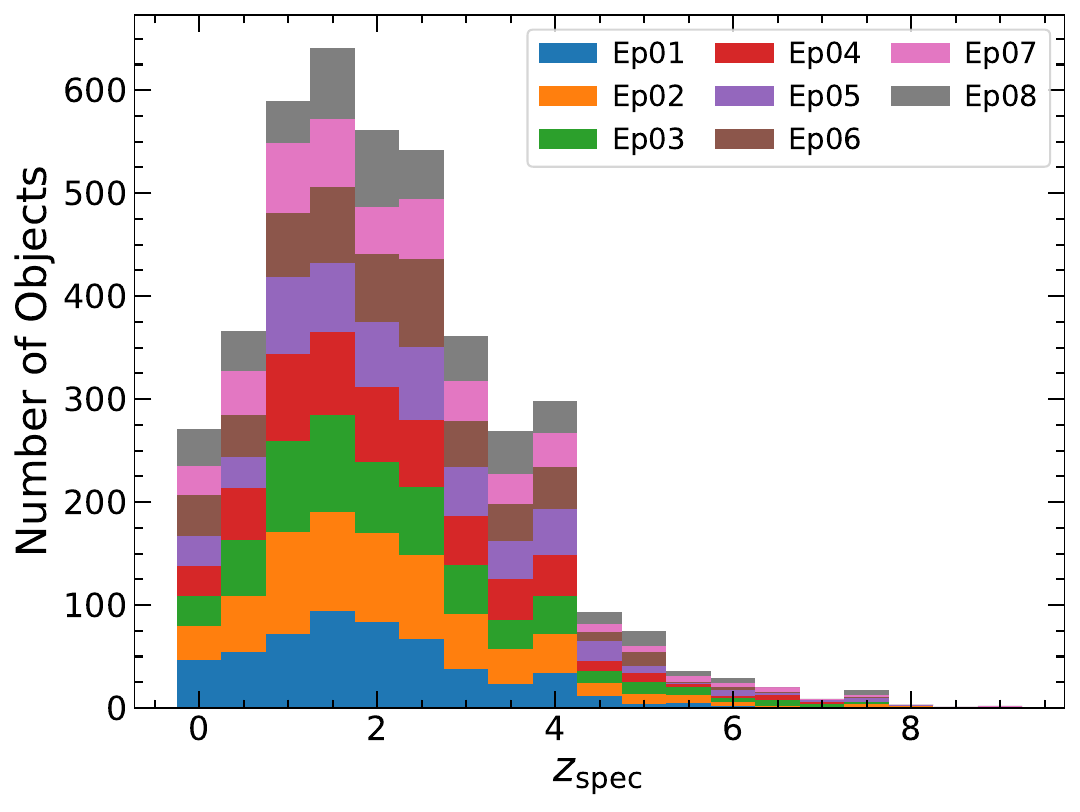}
    \caption{\textbf{Left:} Cumulative number counts of targets observed with NIRSpec MOS PRISM spectroscopy as a function of number of NEXUS-Deep epochs. Magenta, blue, red, green, orange, and brown lines represent Class 0--5, respectively, while the black line represents all source classes combined. \textbf{Middle:} Visually-vetted spectroscopic redshifts ($z_{\rm spec}$) versus F444W magnitude for Deep Epochs 1--4 combined, with magenta, blue, red, green, orange, and brown indicating Class 0, 1, 2, 3, 4, and 5 targets, respectively. Note that transients (class=0.1 and 0.2) and objects with negative F444W fluxes are not shown here. \textbf{Right:} Histogram of $z_{\rm spec}$ for Deep epochs. Colors represent different epochs with correspondence indicated in the legend. }
    \label{fig:deep_cumsum}
\end{figure*}

\begin{itemize}
    \item \textbf{Class 0: transients and variables.} These objects are selected using difference imaging techniques as described in \citet{LeiHu2024} in the F200W and F444W filters observed at different epochs, including those between the first Wide epoch and subsequent Deep epochs, and between individual Deep epochs. This time-domain class started from Deep epoch2, with targets selected using difference imaging between the reference images (Wide epoch 1) and the first Deep epoch. This class further has four subclasses, with Class=0.1 indicating definitive transients, Class=0.2 indicating possible transients, Class=0.3 indicating nuclear variables, and Class=0.4 indicating potential host galaxies associated with a ``hostless'' transient within 2 arcsec. {Here, we consider a transient ``hostless'' if its projected distance to the nearest galaxy is greater than 1 arcsec. Starting from NEXUS-Deep epoch 4, we assign a weight of 40, 10, and 1 for Class=0.1 transients with F200W or F444W$<26$,  26$<$ F200W or F444W $<$ 27, and 27$<$ F200W or F444W $<$ 28; a weight of 4 for Class=0.2 transients; a weight of 20 for Class=0.3 objects, and a weight of 20 or 10 for Class=0.4 objects associated with Class=0.1 or Class=0.2 ``hostless'' transients. We boost weights for ``hostless'' transients and potential high-$z$ transients. For NEXUS-Deep epochs 2 and 3, the weights were generally higher to facilitate more efficient spectroscopic identification of these objects. The exact weight used for target selection can be found by retrieving the APT files from STScI.}

    \item \textbf{Class 1: broad-line active galactic nuclei (AGNs) and little red dots (confirmed or candidates).} These objects are targeted if they satisfy either of the following two criteria: 1. have broad emission lines presented in the optical spectra from DESI DR1 \citep{desi_dr1} or NIR spectra from NIRCam WFSS; 2. meet the criteria used for photometric little red dots (LRD) selection with both ``V-shape'' spectral slopes and compact sizes in the F444W filter as described in \citet{nexus_lrd}. Class 1 objects will remain in the target pool for future Deep epochs to enable variability studies, unless a prior PRISM spectrum invalidates a photometric LRD candidate (e.g., brown dwarfs masquerading photometric LRDs). The initial target list of this class includes 50 sources with weight=20. 

    \item \textbf{Class 2: faint high-$z$ galaxy candidates.} These objects have 26$<$F444W$<$27.5 mag, F200W$<$27 mag, and \zphot$>$8. We include all the compact sources (half-light radius, $R_e<0\farcs15$) and 40 randomly-drawn extended ($R_e>0\farcs15$) sources, {where $R_e$ is measured in F444W}. The initial target list of this class includes 141 sources with weight=20. 

    \item \textbf{Class 3: luminous quasar candidates.} These are relatively bright (F444W$<$24 mag) objects that satisfy the quasar selection criterion in the $z-$F356W (as a substitute for WISE W1) and $g-z$ color diagram as described in \citet{QSO_selection} and have a bright nucleus or barely resolved bulge after visual inspection. {This class does not include objects with confirmed broad emission lines (i.e., Class 1 targets).} {These sources are prioritized as candidates for AGN variability and reverberation mapping studies.} The initial target list of this class includes 126 sources with weight=10. 

    \item \textbf{Class 4: general bright targets.} These objects have F444W$<$26 mag and constitute the bulk of our sample. The initial target list of this class includes 9204 objects with a weight=2 for Deep epoch 1. We increase the weight to 4 starting from Deep epoch 2.

    \item \textbf{Class 5: filler targets.} These are faint objects with F444W fainter than 26 mag but brighter than 27 mag in either F200W or F444W. Given our moderate exposure time of $\sim$20 min, we can still detect continuum emission at a signal-to-noise ratio (SNR) per pixel of $\gtrsim$3 at 2 \micron\ and capture strong emission lines. We did not include these filler targets in Deep epoch 1. Starting from Deep epoch 2, we set their weight to 1 and include them in the MSA target list.

    \item \textbf{Class 6: excluded targets.} These objects are either too faint (F200W$>$27 mag and F444W$>$27) or very bright stars (F200W$<$20 or F444W$<$20; usually saturated) or spurious sources such as spikes, bad pixels, and other artifacts. 
\end{itemize}

Class 0--3 sources are regarded as ``primary'' targets and Class 4 and 5 sources are regarded as ``filler'' targets. For each Deep epoch, we design the MSA masks by maximizing the high-priority primary targets using weights. The exact percentage of each target class receiving a MSA slit varies across epochs, but typically reaches 6--10\%. {We cycle through each successive Deep epoch to improve the targeting efficiency for our high-priority classes. }

After execution of each Deep epoch observation, we examine the spectra of our allocated MSA targets. We move confirmed broad-line emitters to Class 1 for future spectroscopic monitoring. Other observed targets will be excluded from the target pool (weight set to 0) unless the spectrum is unusable due to contamination from very bright nearby sources, large spectral gaps due to detector gaps, flat-fielding issues, or promising high-$z$ targets that require increased exposure times. Transient and variable objects discovered from new NIRCam imaging are added to the MSA target pool for future MSA epochs. 

\subsection{General caveats}


As mentioned in Section~\ref{sec:msa_gen}, each NEXUS-Deep NIRSpec MOS pointing has only two exposures in a two shutter slitlet nodding pattern. This may lead to compromised performance of outlier rejection algorithm and produce spurious emission lines in rare cases. 

During the MSA planning process, we allow spectral gaps in epochs 1--3 and request no spectral gaps for our primary targets in later epochs. While this maximizes the multiplexing capability of MSA, it unavoidably results in a small fraction ($\sim$10\%) of Class$\leq$4 objects showing spectral gaps greater than 30\% of the total wavelength coverage. We add those targets back to the planning pool so that they are eligible for future MSA target selection.

We also allow the opening of shutters containing multiple sources. Primary and filler sources with nearby neighbors that occupy the same shutter will still be able to receive a slit at the risk of overlapping spectra from nearby neighboring ``contaminants''. Only $\sim$10 primary or filler targets per epoch fail the spectra extraction process due to this complication.

\begin{deluxetable}{ccc}[ht]
\tablecaption{Column Description of MSA Target Summary Table\label{tab:MSA_target}}
\tablewidth{0pt}
\tablehead{
\colhead{Column} & \colhead{Units} & \colhead{Description}
}
\startdata
MSAID &  & Target ID\\
RA &  degrees & R.A. (J2000)\\
DEC &  degrees & Decl. (J2000)\\
Source Type &   & Type of the target\\
Priority Class &  & Target Priority Class\\
weight &  & Weight for MSA planning\\
F200W & $\mu$Jy & F200W flux\\
F200W\_e & $\mu$Jy & F200W flux error\\
F444W & $\mu$Jy & F444W flux\\
F444W\_e & $\mu$Jy & F444W flux error\\
z\_VI &  & Visually inspected redshift\\
z\_VI\_conf &  & Confidence level of z\_VI\\
\enddata
\tablecomments{``MSAID'' is a different designation from our fiducial NEXUS ID (NID). ``Source Type'' includes ``Primary'' for Class=0--3 sources, ``Filler'' for Class=4 and 5 sources, and ``Contaminant'' for sources contaminating ``Primary'' and ``Filler'' targets. F200W and F444W are total fluxes measured on the NIRCam images obtained in the same Deep epoch as the NIRSpec MOS observations, using a 0\farcs3-diameter circular aperture and corrected to total flux via aperture corrections. The flux measurement procedure follows that described in \citet{nexus_edr}. Initial spectroscopic redshifts are derived using the \texttt{fit\_redshift} function in \texttt{msaexp}, adopting the \texttt{agn\_blue\_sfhz\_13} template set from EAZY \citep{EAZY}, and are subsequently visually inspected by M.~Zhuang. z\_VI\_conf: 0=no robust z\_VI; 1=ambiguous z\_VI; 2=robust z\_VI.}
\end{deluxetable}

\subsection{Overview of quick release data}


We download the stage 1 data products (rate.fits files) of JWST calibration pipeline from the Mikulski Archive for Space Telescopes (MAST\footnote{\url{https://mast.stsci.edu/}}) and reduce all the NIRSpec MOS PRISM spectra using \texttt{v0.9.12} of the \texttt{msaexp} software \citep{msaexp} paired with the JWST calibration pipeline v1.16.1 and the reference file \texttt{jwst\_1298.pmap}. \texttt{msaexp} performs the JWST stage 2 pipeline \texttt{calwebb\_spec2} and applies additional steps including 1/f noise correction, bias removal, readout noise rescaling using empty parts of the exposure, and modifies the wavelength extraction limit. We subtract the background using the differences of nodded exposures and optimally extract the 1D spectrum from the coadded spectrum.

For each Deep epoch, we release the following Quick-release data products:

\begin{enumerate}

    \item MSA target summary files (columns described in Table~\ref{tab:MSA_target}), one for each epoch, placed under the ``/nirspec/'' subdirectory. It contains MSAID (ID for MSA planning), coordinates, source type, priority class, weight, total F200W and F444W fluxes and their associated errors, and visually inspected redshift and its confidence level. 
    
    \item {NIRCam imaging mosaics and their corresponding uncertainty maps in F200W and F444W filters under the ``/nircam/'' subdirectory. These images were reduced following the procedures described in \citet{nexus_edr} and are sampled at a pixel scale of 30 mas. For the F444W filter, we provide an additional version at 60 mas pixel scale to mitigate sub-sampling artifacts and ``pixelated'' PSF shapes resulting from the limited two-point dither pattern. A more comprehensive description of the reduction process will be presented in the forthcoming first data release (DR1) paper (Zhuang et al., in prep).}
    
    \item Quick-reduction MSA spectra. The spectral fits files are placed under ``/nirspec/fits/'' and the quick-look spectral plots are under ``/nirspec/plots/''.

\end{enumerate}

New epoch data will be released approximately 2 months after observation. All QR data products can be accessed from the NEXUS data repository: \\
https://ariel.astro.illinois.edu/nexus/qdr/

{Figure~\ref{fig:deep_cumsum} shows the cumulative number counts of targets in each class, visually-vetted redshifts versus F444W mag, and the histogram of visually-vetted redshifts of the targets for all epochs taken so far. We will update this figure each time we release the new Deep epoch. {One of the main goals of NEXUS is to obtain reliable spectroscopic redshifts for a complete sample with F444W brighter than 26 mag. Our visual inspection suggests that approximately 60--70\% of objects show significant emission line features. The remaining objects are mostly continuum-dominated systems with a small fraction affected by spectral gaps. {In each MSA epoch, approximately 550 objects have robust spectroscopic redshift after visual inspection.}

{The NEXUS Deep-tier data enable a broad range of scientific investigations spanning stellar and galaxy evolution, black hole growth, and time-domain astrophysics. The multi-epoch NIRSpec spectroscopy and coordinated imaging provide a powerful dataset to identify and characterize AGNs, including LRDs, across a wide range of luminosities and obscuration levels, and to investigate the physical properties of both star-forming and quiescent galaxies over a broad parameter space. The uniform availability of spectroscopic redshifts for an unbiased, flux-complete galaxy sample (F444W$\lesssim$26 mag) will substantially improve spectroscopic training sets for photometric redshift estimation. The regular cadenced observations further open a new window for high-redshift variability studies, enabling the discovery of transients and the characterization of both continuum and emission-line variability in AGNs.
}



\section{Detailed Deep epoch notes}\label{sec:deep_notes}

\subsection{Deep1: 2025/06/01}

\begin{figure*}[ht]
    \centering
    \includegraphics[width=0.5\linewidth]{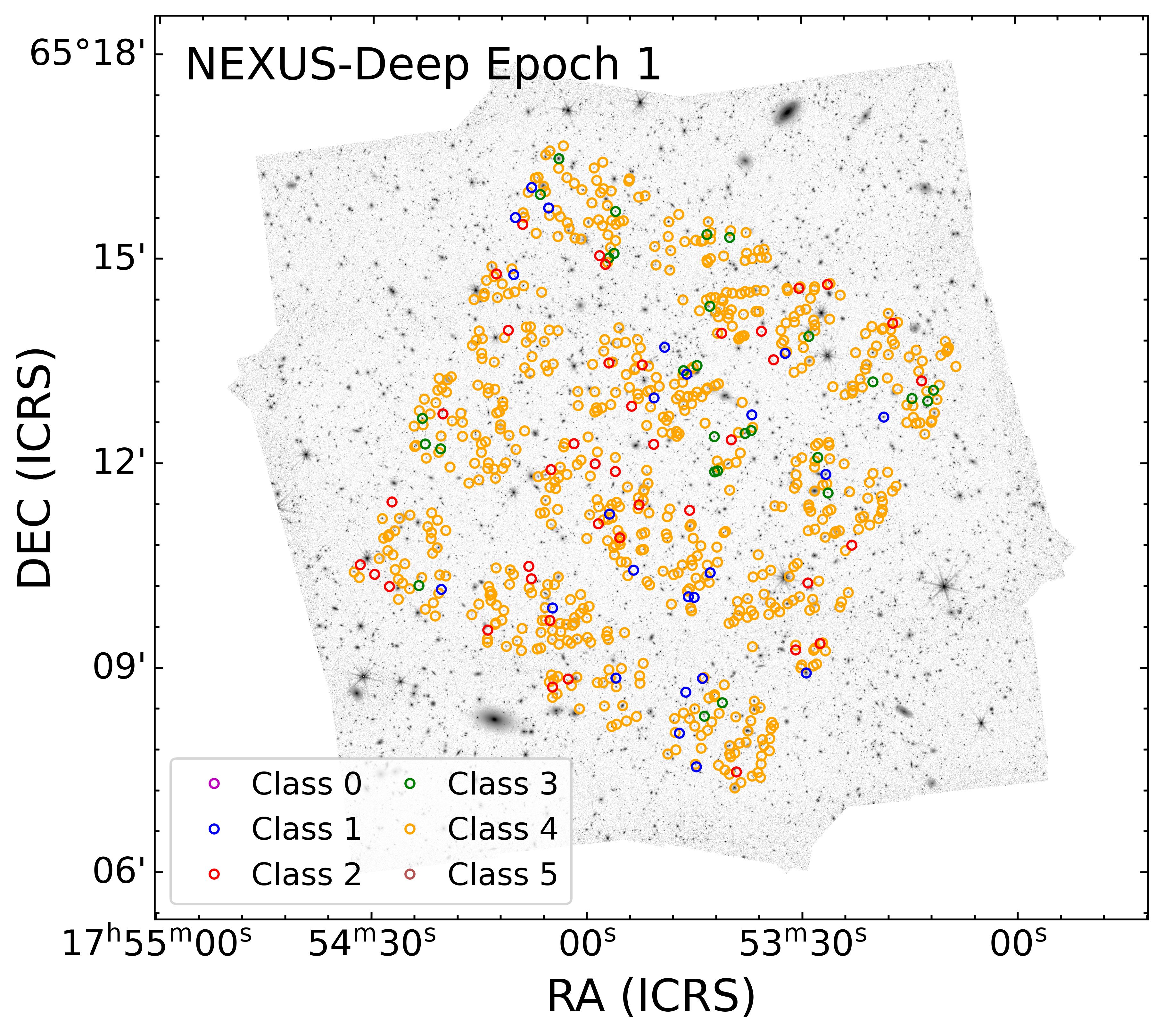}
    \includegraphics[width=0.46\linewidth]{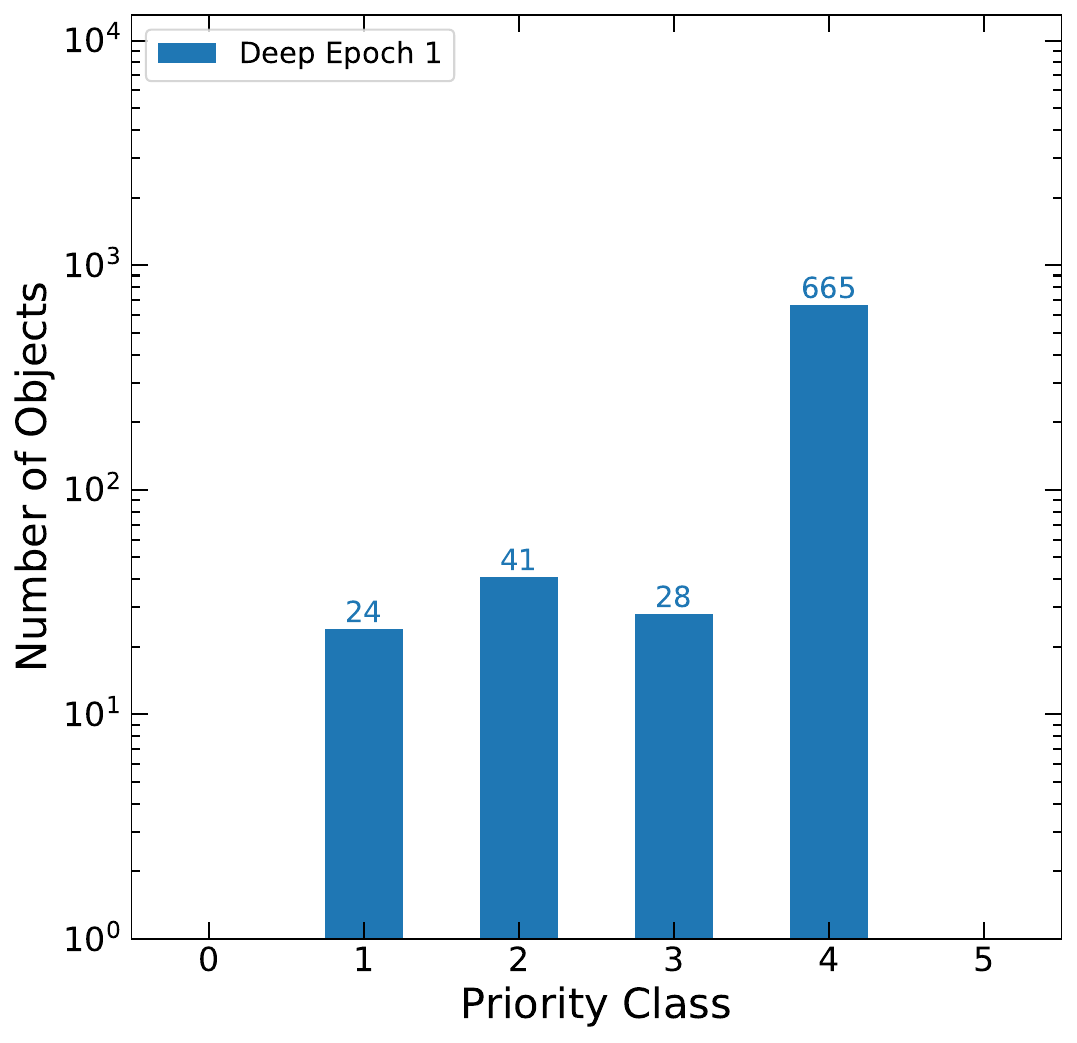}
    \includegraphics[width=0.46\linewidth]{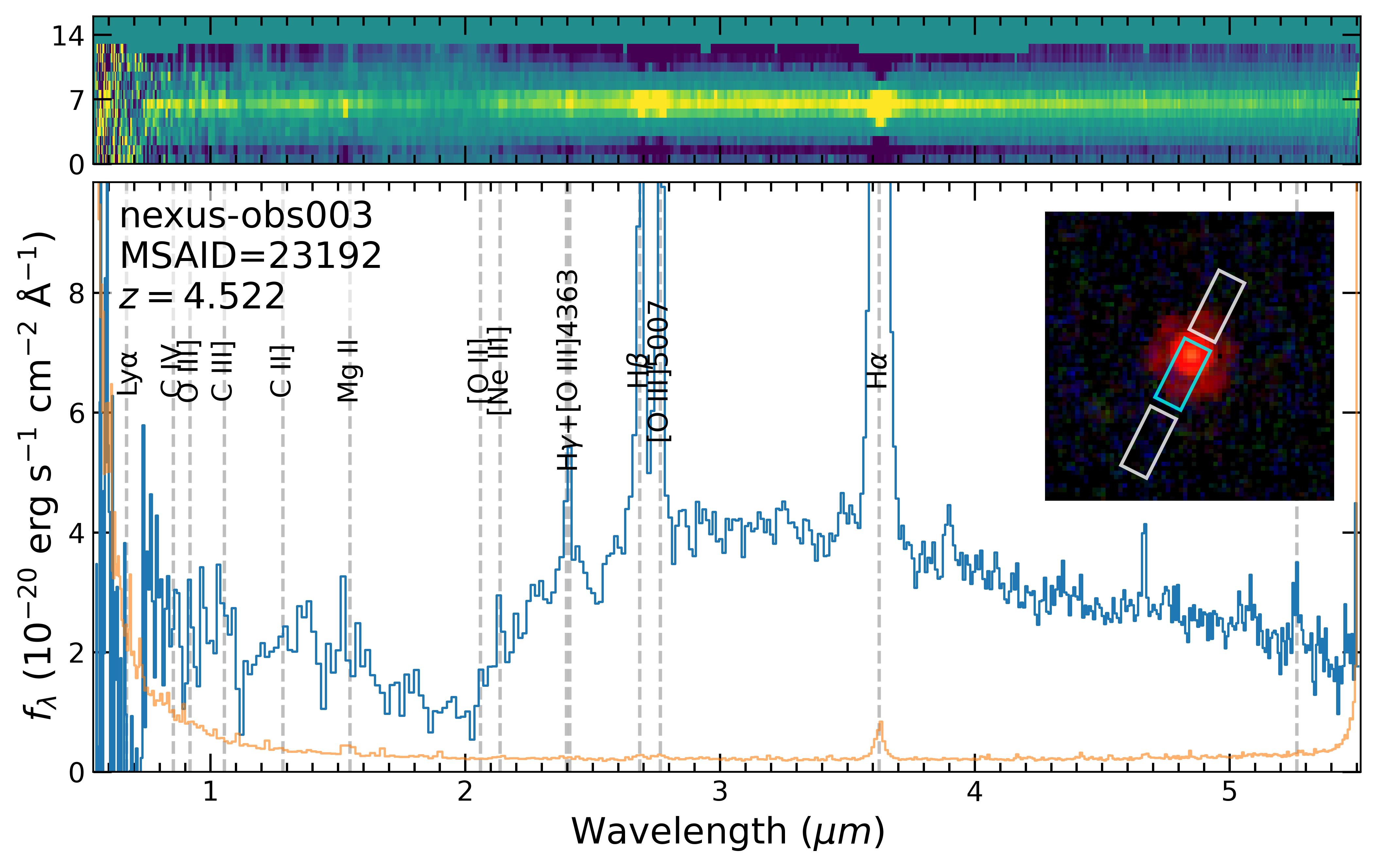}
    \includegraphics[width=0.46\linewidth]{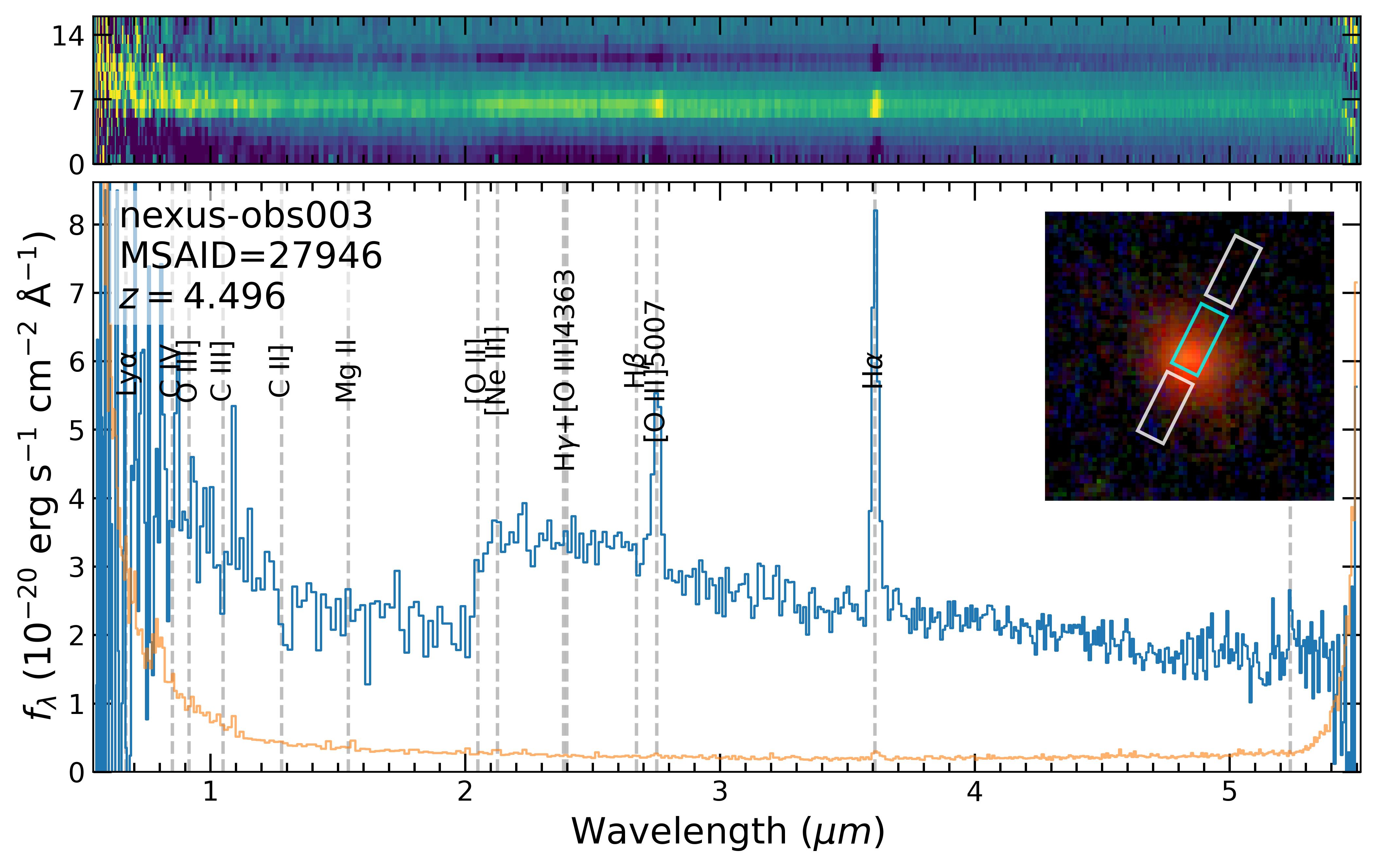}
    \caption{\textbf{Top: } Targets layout (left) and distribution of target classes (right). Class 0, 1, 2, 3, 4, and 5 are indicated by magenta, blue, red, green, orange, and brown open circles, with stacked F444W image shown in the background. Blue and black bars indicate the number of objects  in current epoch and all observed epochs (including the current epoch). \textbf{Bottom: } Two-dimensional (2D) and one-dimensional (1D) spectra of two noteworthy targets from this epoch. The inset shows the color composite image (RGB: F444W–F200W–F090W) together with the slitlet configuration. Prominent emission lines are labeled and marked with vertical dashed lines.}
    \label{fig:deep01}
\end{figure*}

{As mentioned in Section~\ref{sec:deep}, each NIRSpec MSA epoch consists of four pointings. The initial pointing centers are distributed along a circle with a radius of 2.5 arcmin, centered on the NEXUS field center (R.A.=17:53:51 Decl.=+65:11:57). These positions are optimized based on the aperture position angle (PA) to ensure that the four pointings tile the region seamlessly without significant
gaps. Final pointing centers are determined by maximizing the sum of weight of objects that could be accommodated in MSA shutters. The initial target pool, based on the NEXUS EDR catalog \citep{nexus_edr}, includes all sources within a 5 arcmin-radius circle of the NEXUS center.}

{The top panels in Figure~\ref{fig:deep01} show the targets layout in Deep Epoch 1 and the histogram of targets observed so far per priority class. In Deep Epoch 1, we observed 24 class=1 objects, 41 class=2 objects, 28 class=3 objects, 665 class=4 objects, 0 class=5 objects, with 758 objects in total. 21 contaminants also fell within open slitlets.} 

{In Deep Epoch 1, we highlight two noteworthy sources in Fig.~\ref{fig:deep01}: a $z=4.522$ little-red-dot (LRD; MSAID=23192) and a $z=4.496$ post-starburst (MSAID=27946). Source 23192 exhibits a characteristic ``V''-shape spectrum with a prominent broad \Ha\ line, consistent with LRDs identified in other JWST programs \citep[e.g.,][]{Greene+2024, Taylor+2024}. This object was also classified as an LRD by \citet{nexus_lrd} based on its morphology, spectral shape, and broad \Ha\ emission in NIRCam WFSS. Source 27946 displays clear post-starburst features, including strong Balmer break and stellar emission dominated by A-type stars. Rather than the deep absorption features typical of quiescent galaxies, it shows strong \OIII\ and \Ha+\NII\ emission lines, suggesting ongoing AGN activity. It closely resembles the massive quiescent galaxy GS9209 at $z=4.658$ reported in \citet{2023Natur.619..716C}, and it likely represents a descendant of high-redshift quasars and extreme starbursts.}




\subsection{Deep2: 2025/07/30}
{\textbf{Summary of changes}: 1. To mitigate the $\sim$20 arcsec gaps present in the initial NEXUS EDR NIRCam mosaics \citep{nexus_edr}, we improved our source detection by stacking the EDR data with Deep Epoch 1 NIRCam mosaics. This combined detection map was used to generate an updated MSAID catalog, significantly increasing tiling efficiency and target completeness. 
2. Starting from Deep Epoch 2, the target pool was expanded to include transients and variable sources (classified as class=0), following the selection criteria detailed in Section~\ref{sec:msa_gen}. }

{The top panels in Figure~\ref{fig:deep02} show the targets layout of Deep Epoch 2 and histogram of targets observed so far as a function of priority class. We observed 12 class=0 objects, 25 class=1 objects, 39 class=2 objects, 30 class=3 objects, 681 class=4 objects, 184 class=5 objects, with 971 objects in total. 35 contaminants also fell within open slitlets.}

{For Deep Epoch 2, we highlight a T-type brown dwarf (MSAID=13038) and a $z=2.658$ quasar (MSAID=20616). Source 13038 exhibits characteristic T-dwarf spectral features, including deep $\rm H_2O$ bands at $\sim$1.4 and 1.9 $\mu$m, strong $\rm CH_4$ absorption at $\sim$1.6 and 2.2 $\mu$m, and a prominent 3.3 $\mu$m $\rm CH_4$ fundamental band. Source 20616 is a bright quasar with a bolometric luminosity of $\sim8\times10^{45}$ erg s$^{-1}$. Its spectrum shows strong, broad emission lines including \CIV, \CIII, \MgII, \Hb, \Ha, \HeI, and \Pab, typical of other bright AGNs in the NEXUS field. As a primary monitoring target, source 20616 will be observed as many epochs as possible to measure spectral variability. }


\begin{figure*}[ht]
    \centering
    \includegraphics[width=0.5\linewidth]{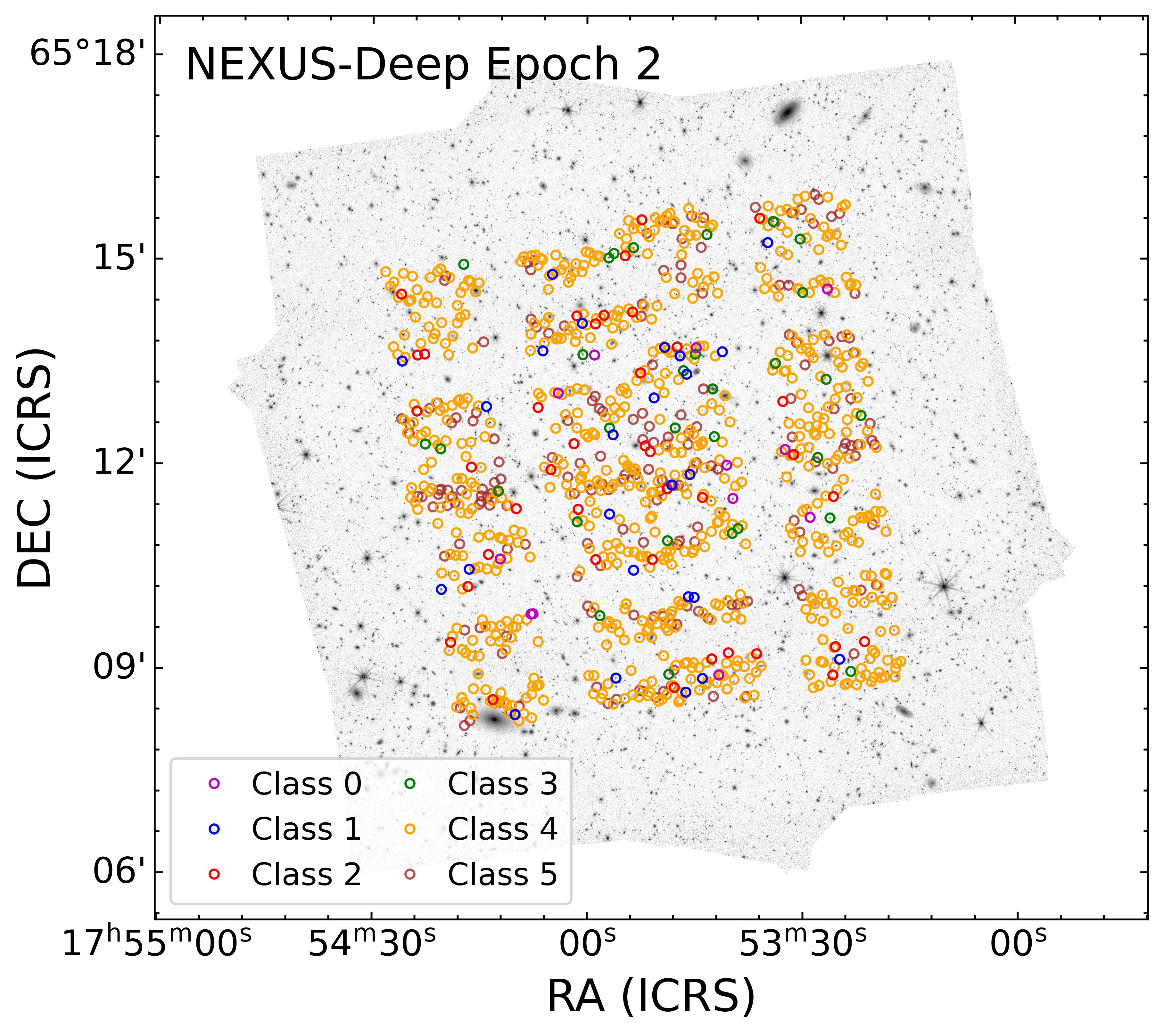}
    \includegraphics[width=0.46\linewidth]{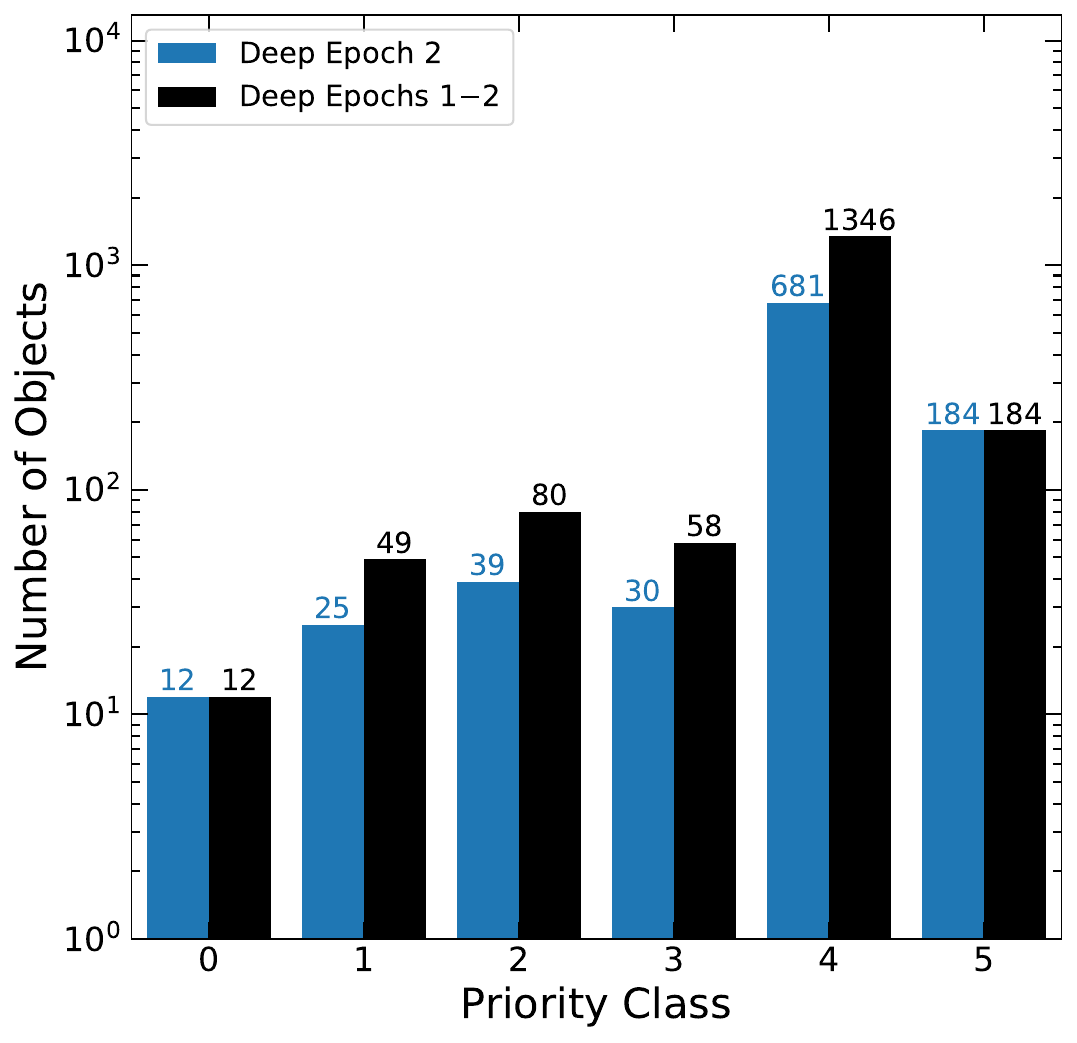}
    \includegraphics[width=0.46\linewidth]{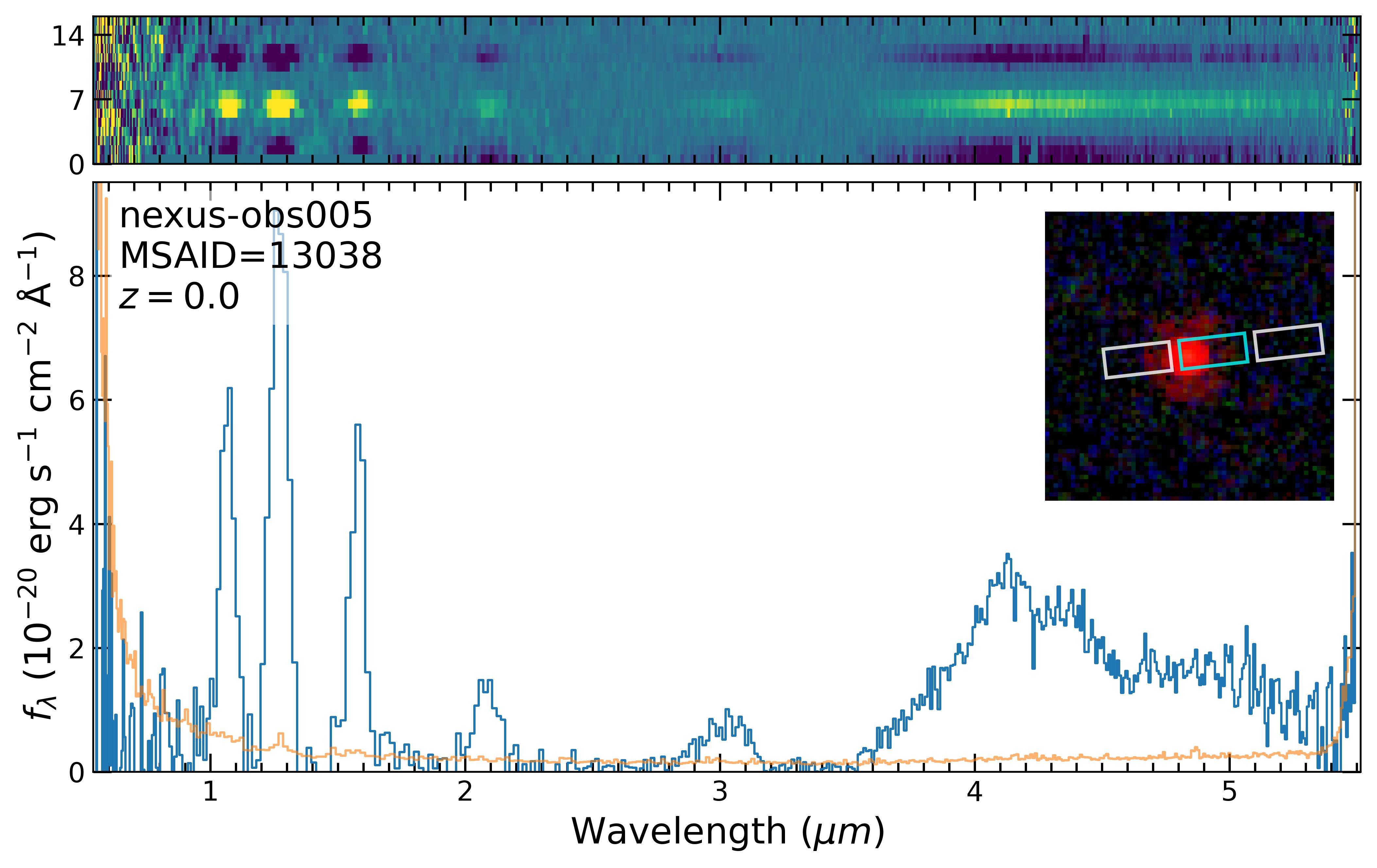}
    \includegraphics[width=0.47\linewidth]{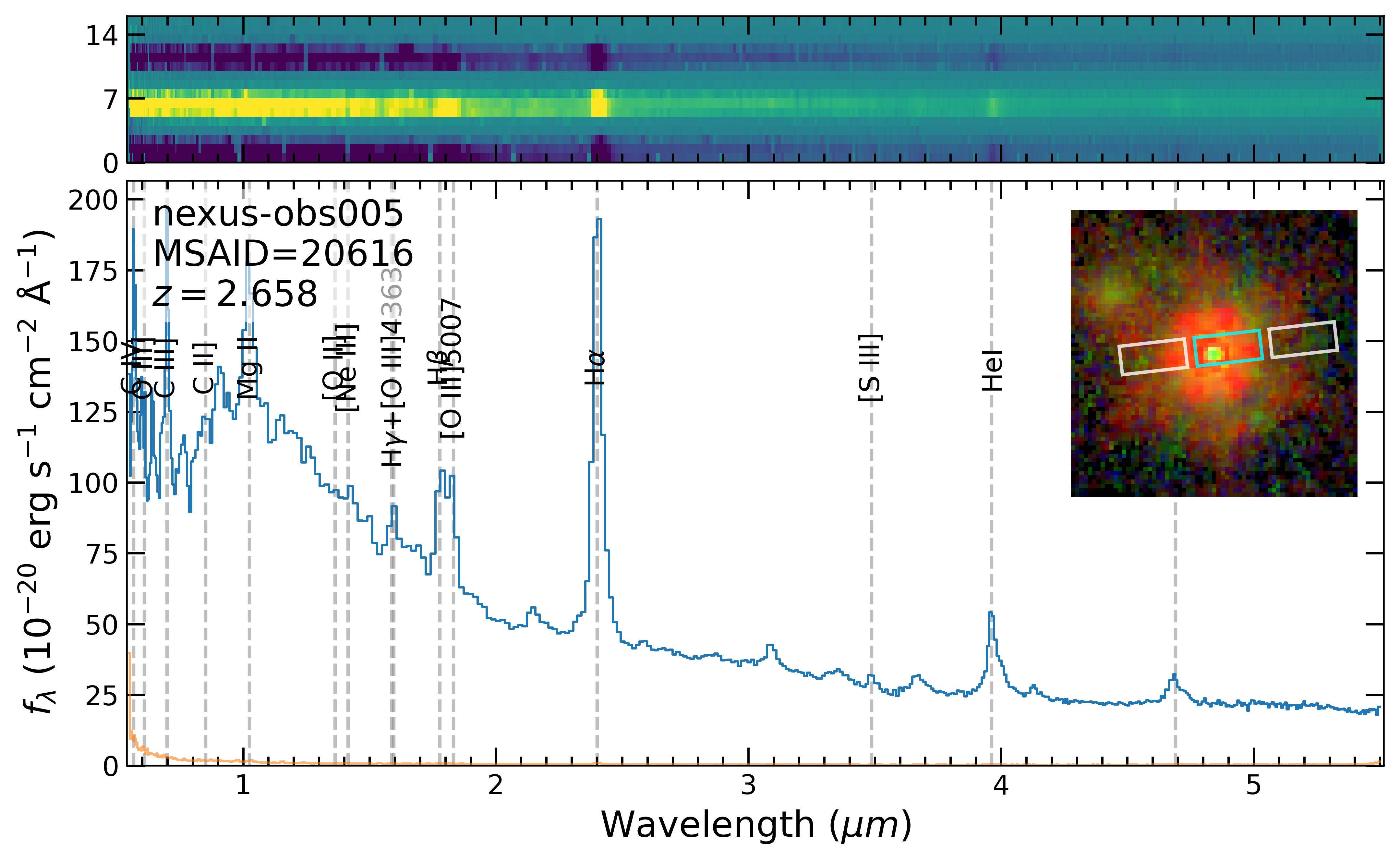}
    
    \caption{Same as Figure~\ref{fig:deep01} but for Deep Epoch 2.}
    \label{fig:deep02}
\end{figure*}

\subsection{Deep3: 2025/09/28}

{\textbf{Summary of changes}: 
1. We have implemented a consistent re-reduction of all previous NIRCam imaging data, including NEXUS Wide Epoch 1 and Deep Epochs 1 and 2. This processing achieved better astrometric precision, with a root-mean-square error of $\sim$8 mas. We subsequently updated the source catalog using these stacked mosaics. The increased depth of the combined detection map allowed us to augment the target pool with a significant number of newly identified sources.
2. To optimize the multiplexing efficiency of the MSA, particularly for transients and variable sources located near the field boundaries, we increased the target pool radius from $r \leq 5'$ to $r \leq 5.4'$ from the NEXUS center. This adjustment provides greater flexibility for the MSA planning software to accommodate class=0 sources near the edge of the footprint while maintaining a high slit density. 
}

{The top panels in Figure~\ref{fig:deep03} show the targets layout of Deep Epoch 3 and the histogram of targets observed so far as a function of priority class. We observed 12 class=0 objects, 12 class=1 objects, 25 class=2 objects, 9 class=3 objects, 683 class=4 objects, 181 class=5 objects, with 922 objects in total. 55 contaminants also fell within open slitlets.}

{In Deep Epoch 3, we highlight a chance alignment of a $z=1.430$ galaxy (MSAID=5211) with a background object at $z=4.886$, alongside a $z=1.494$ dual AGN candidate (MSAID=25745). Source 5211 is a clumpy, star-forming galaxy that appears to be interacting with a companion to the southeast. Notably, its spectrum reveals two distinct sets of emission lines at $z=1.430$ and at $z=4.886$, confirming the line-of-sight overlap. Source 25745 consists of two bright nuclei with a projected separation of $\sim$2 kpc. Its Deep Epoch 3 spectrum shows a red continuum with prominent \OIII, \Ha-\NII\ complexes, and \HeI\ emission lines. The elevated \OIII/\Hb\ ratio is suggestive of a narrow-line AGN. Although the Deep Epoch 3 slit was centered on the brighter nucleus, data from other epochs cover both nuclei at different PAs. This would allow spatial decomposition of the two nuclei to confirm their dual AGN nature.}

\begin{figure*}[ht]
    \centering
    \includegraphics[width=0.5\linewidth]{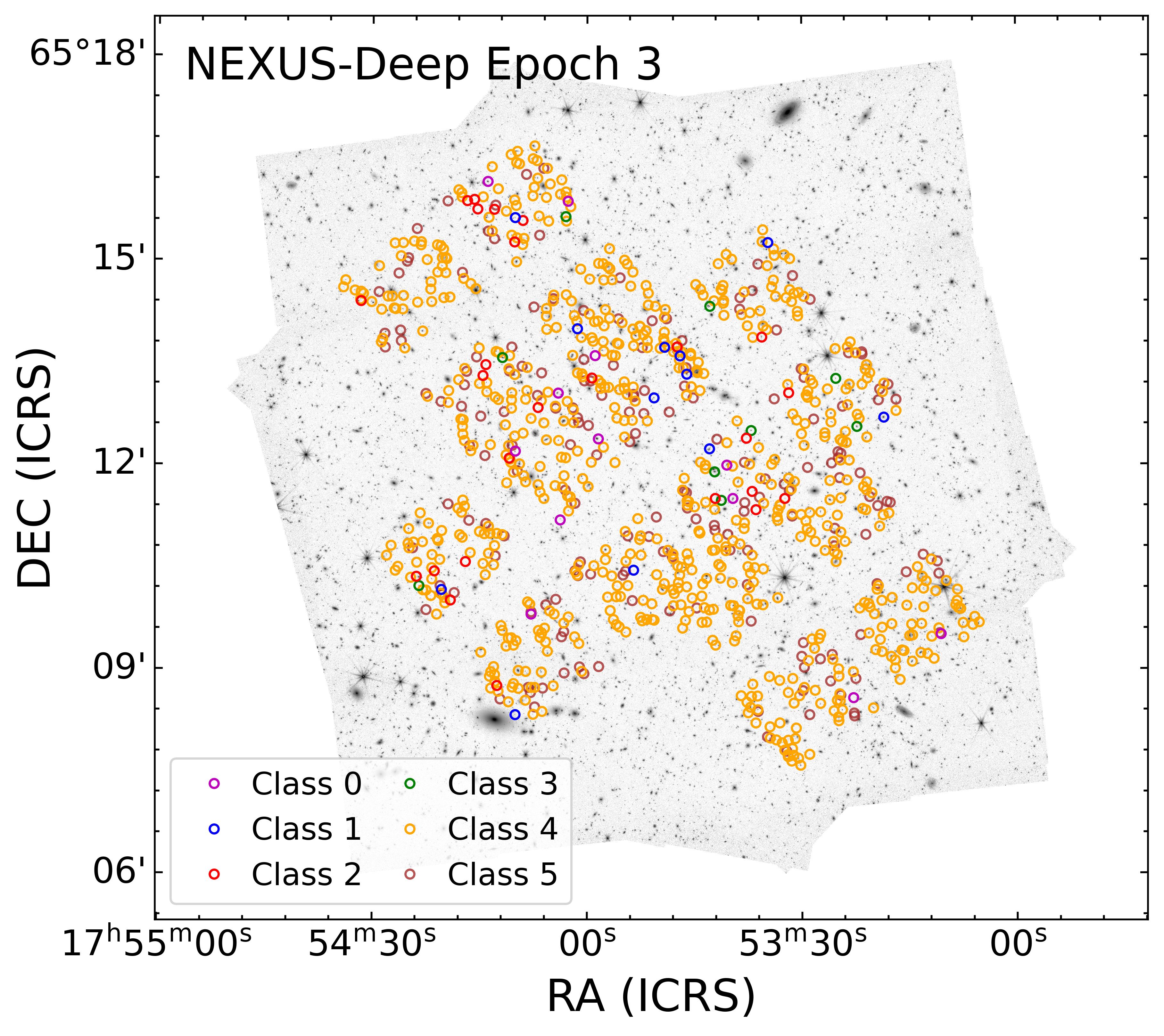}
    \includegraphics[width=0.46\linewidth]{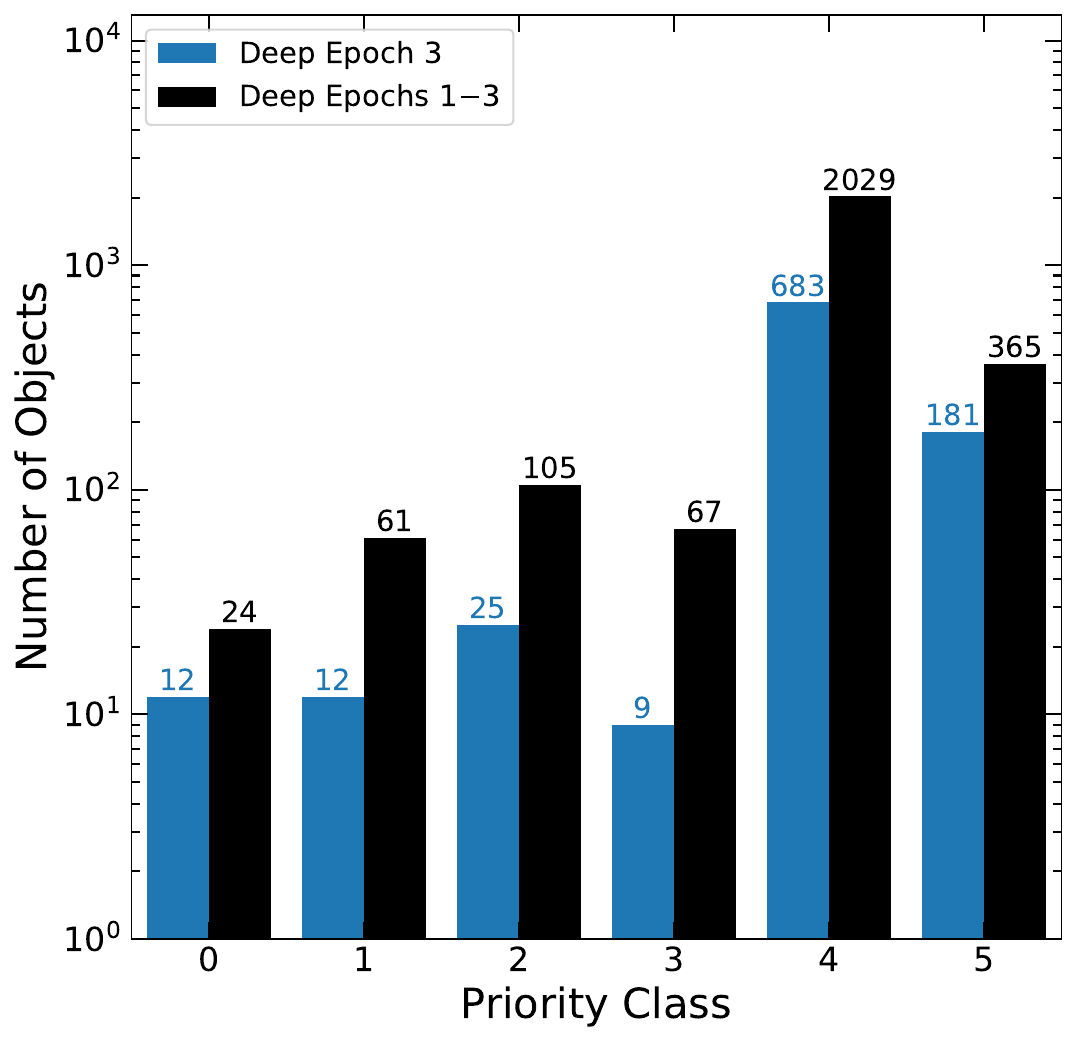}
    \includegraphics[width=0.46\linewidth]{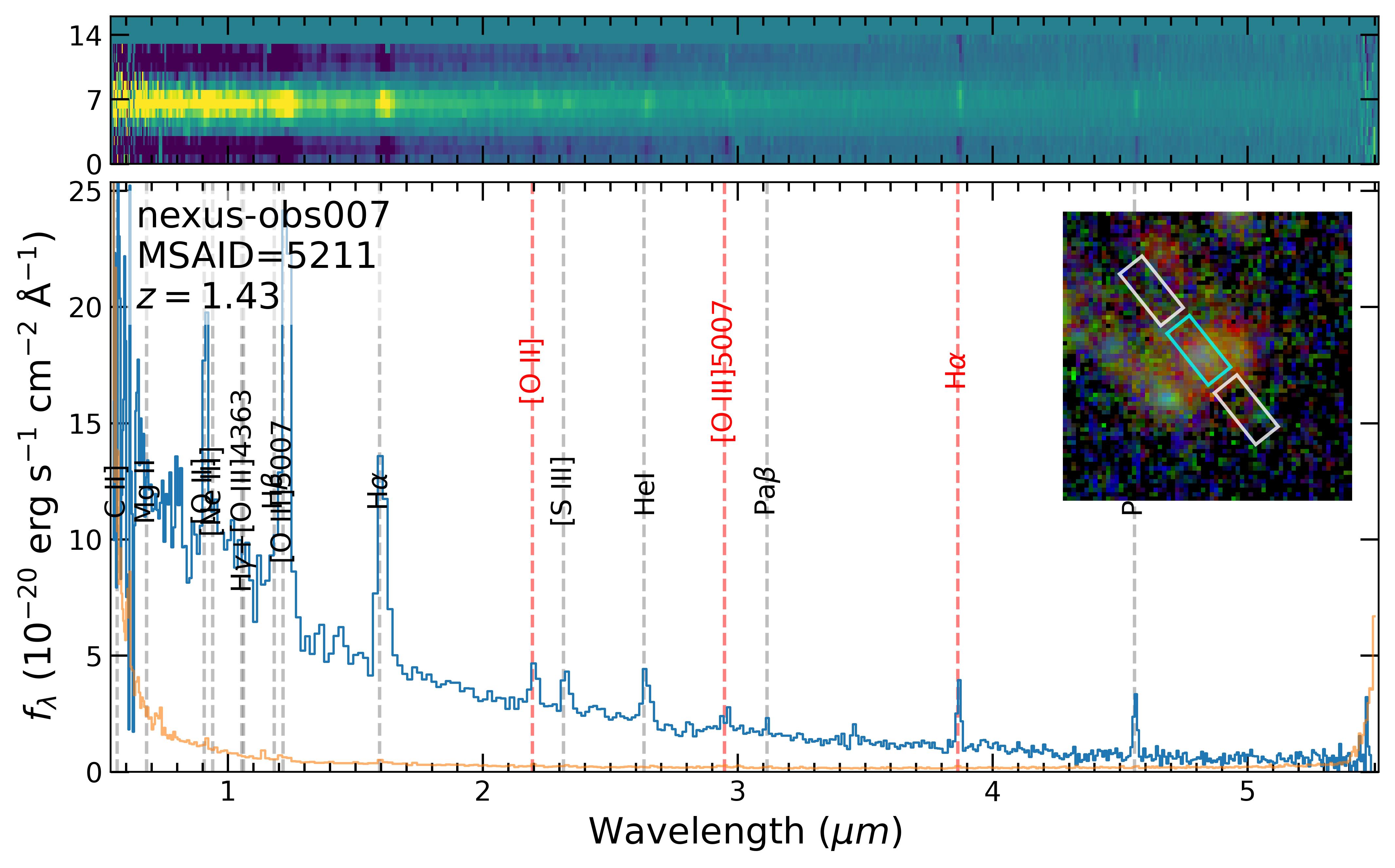}
    \includegraphics[width=0.47\linewidth]{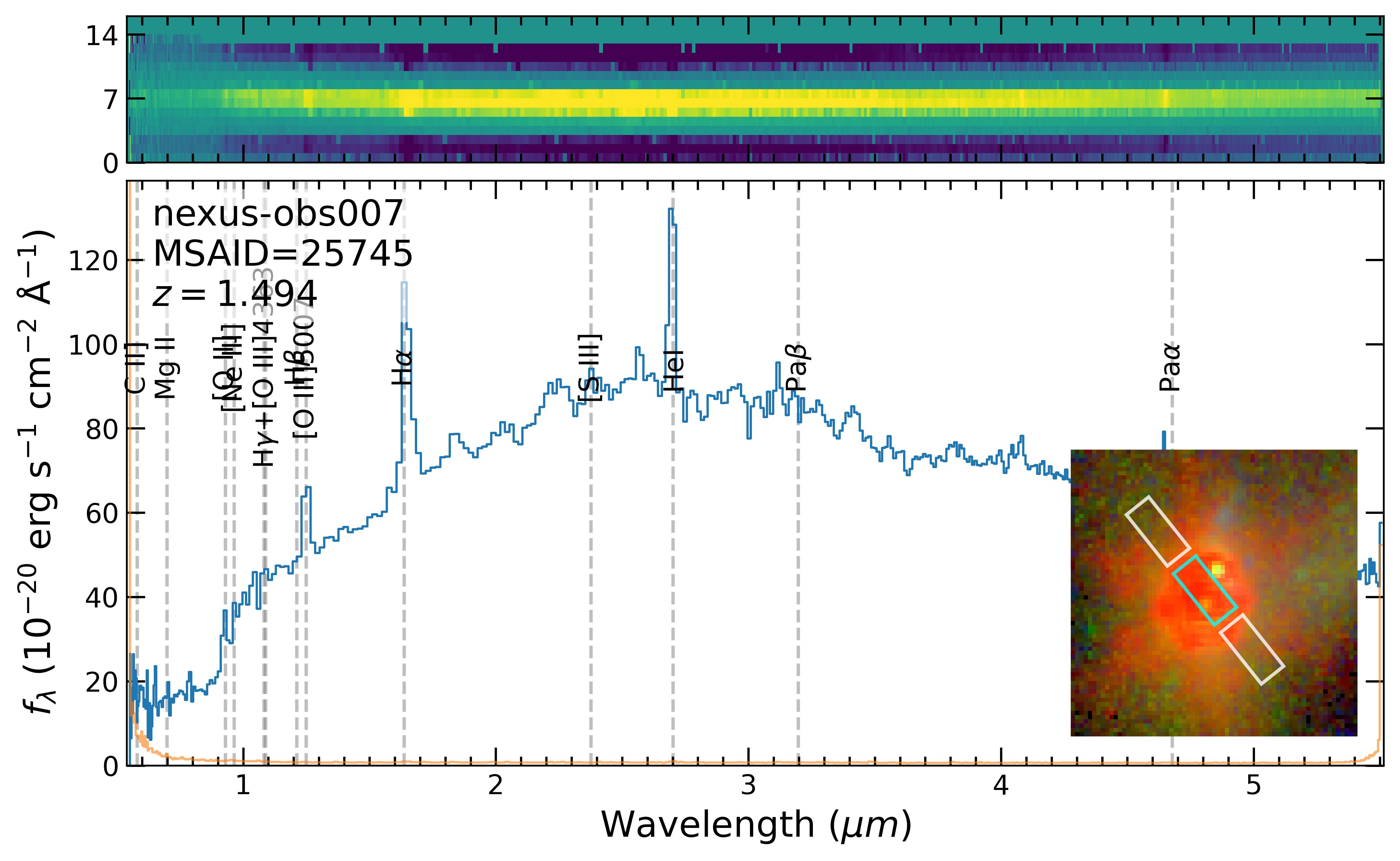}
    \caption{Same as Figure~\ref{fig:deep01}  but for Deep Epoch 3. In the example target shown in the lower-left panel, the red labels and dashed lines indicate emission lines from a background galaxy at $z=4.886$.}
    \label{fig:deep03}
\end{figure*}

\subsection{Deep4: 2025/11/28}

{\textbf{Summary of changes}: We require our primary targets ($\text{class}=0$--$3$) not have spectral gaps to ensure high-fidelity redshift determination and robust subsequent analysis. While this constraint slightly reduces the overall multiplexing efficiency, it is necessary to prevent critical diagnostic features from falling into the physical gaps between NIRSpec detectors.}

{The top panels in Figure~\ref{fig:deep04} show the targets layout of Deep Epoch 4 and the histogram of targets observed so far as a function of priority class. We observed 7 class=0 objects, 14 class=1 objects, 27 class=2 objects, 11 class=3 objects, 658 class=4 objects, 210 class=5 objects, with 927 objects in total. 41 contaminants also fell within open slitlets.}

{In Deep Epoch 4, we highlight a luminous $z=6.623$ galaxy (MSAID=2698) and a heavily reddened galaxy likely at $z=3.829$ (MSAID=6003), in Fig.~\ref{fig:deep04}. Source 2698 is a luminous, low-mass star-forming galaxy at cosmic dawn, characterized by its high equivalent width \OIII\ and \Ha\ emission lines. With an ultraviolet (UV) magnitude of $M_{\rm UV}=-20.74$ mag, it straddles the knee ($L^*$) of UV luminosity at $z=6-7$. Source 6003 is a very red and extended object with extremely red color F200W$-$F444W=3 mag and F115W$-$F444W=4.5 mag. Only a single strong emission line is detected at $\sim$3.2 $\mu$m, which we tentatively identify as \Ha\ at $z=3.829$. This suggests that source 6003 is an extremely dusty star-forming galaxy. However, multi-wavelength data, particularly in sub-millimeter wavelength, are required to fully understand its nature.}

\begin{figure*}[ht]
    \centering
    \includegraphics[width=0.5\linewidth]{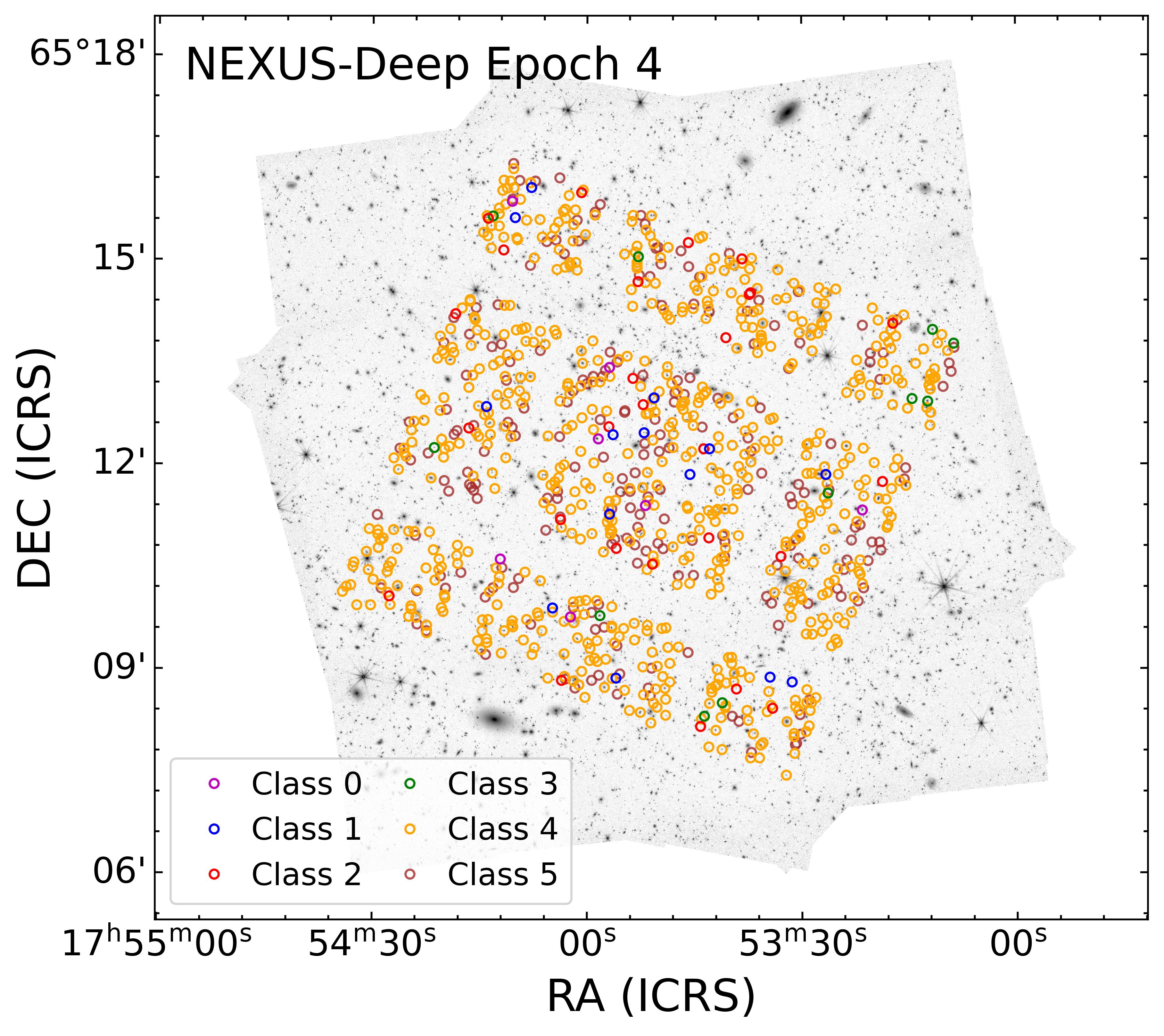}
    \includegraphics[width=0.46\linewidth]{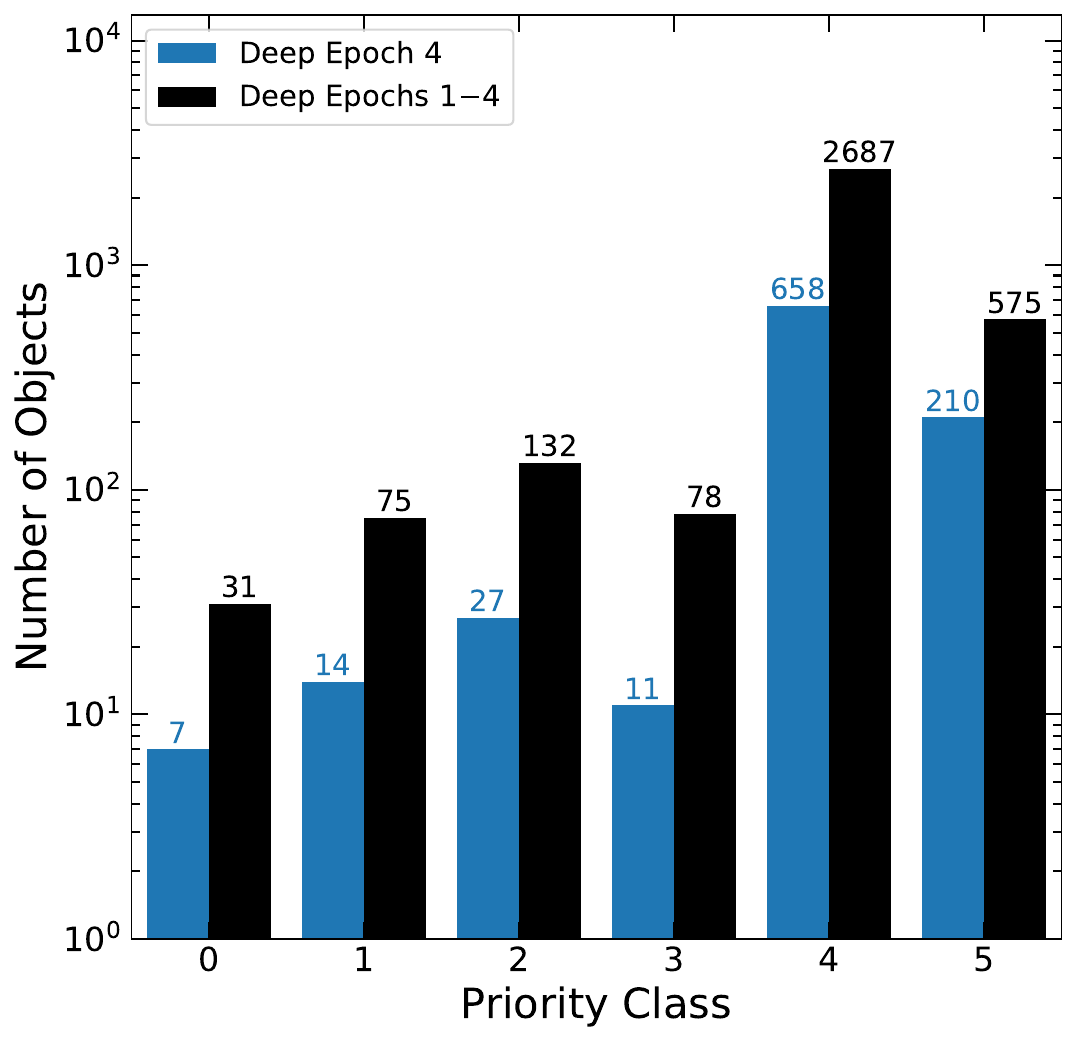}
    \includegraphics[width=0.46\linewidth]{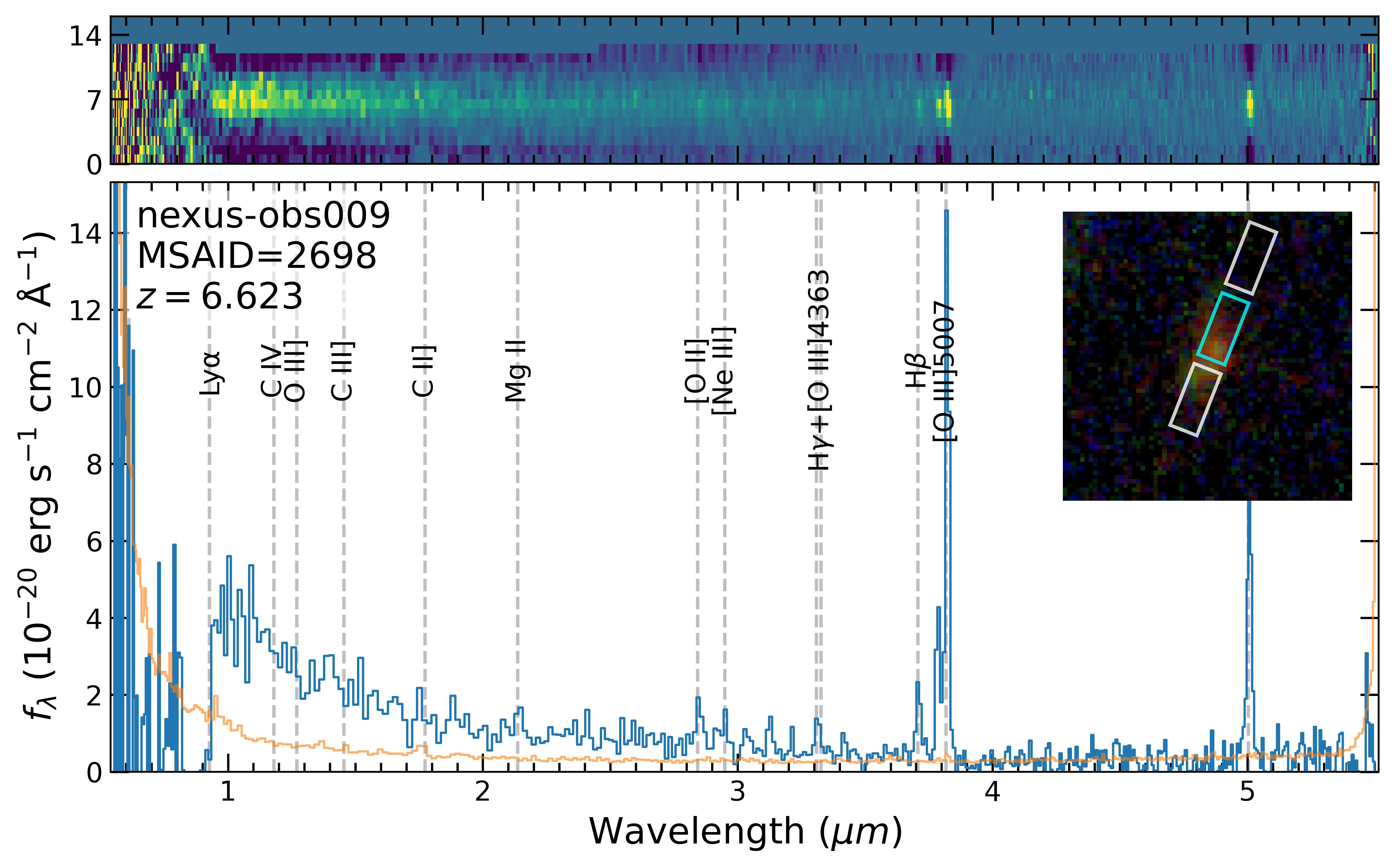}
    \includegraphics[width=0.46\linewidth]{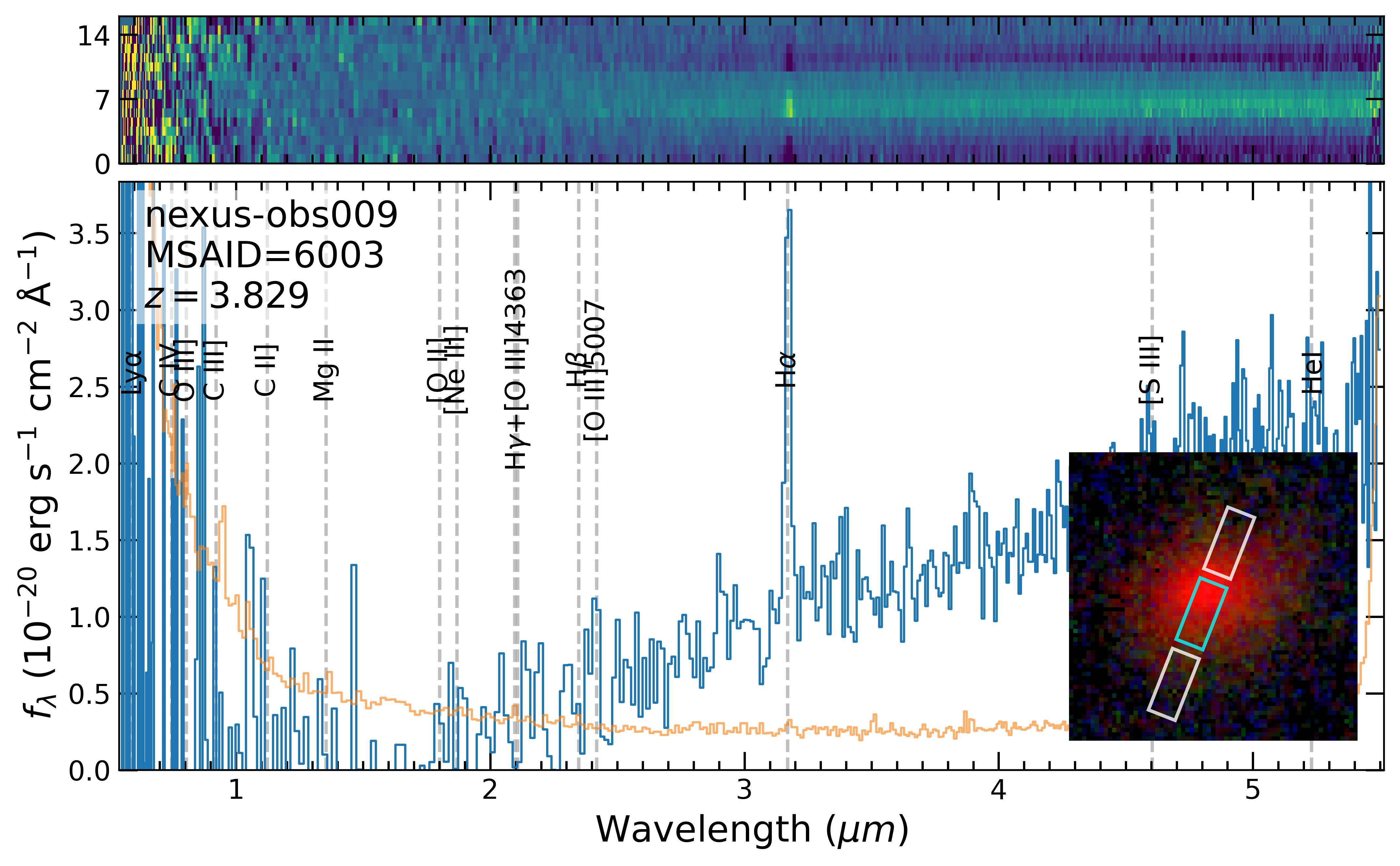}
    \caption{Same as Figure~\ref{fig:deep01} but for Deep Epoch 4.}
    \label{fig:deep04}
\end{figure*}

\subsection{Deep5: 2026/1/28}

{\textbf{Summary of changes}: 1. Previously observed Class=4 objects with spectral gaps exceeding 30\% over the wavelength range of 0.6--5.3 \micron\ were returned to the target pool to provide a second opportunity for obtaining a more complete spectrum. 2. Previous primary targets without a final extracted spectrum, mainly due to overlapping spectra from contaminants, were also added back to the target pool.}

The top panels in Figure~\ref{fig:deep05} show the targets layout of Deep Epoch 5 and the histogram of targets observed so far for different priority classes. We observed 9 class=0 objects, 18 class=1 objects, 15 class=2 objects, 14 class=3 objects, 612 class=4 objects, 241 class=5 objects, with 909 objects in total. 45 contaminants also fell within open slitlets.

In Deep Epoch 5, we highlight the host galaxies of two transients at $z=1.946$ (MSAID=36677) and $z=2.772$ (MSAID=36693). Both galaxies exhibit prominent nebular emission lines, including \Hb, \OIII$\lambda5007$, \Ha, \SIII$\lambda9532$, \HeI$\lambda10830$, and \Pab, indicating vigorous ongoing star formation. Unfortunately, the transient associated with source 36677 faded below the detection limit at the time of Deep Epoch 5, while the spectrum of source 36693 is dominated by host-galaxy emission, making spectroscopic characterization of the transient itself nearly impossible. Nevertheless, the regular cadence of the NEXUS Deep Epoch observations enables a systematic study of transient number densities and their cosmic evolution through the properties of their host galaxies. A detailed study of transients discovered in the first year of NEXUS observations (first six Deep Epochs) is presented in \citet{nexus_transients}.

\begin{figure*}[ht]
    \centering
    \includegraphics[width=0.5\linewidth]{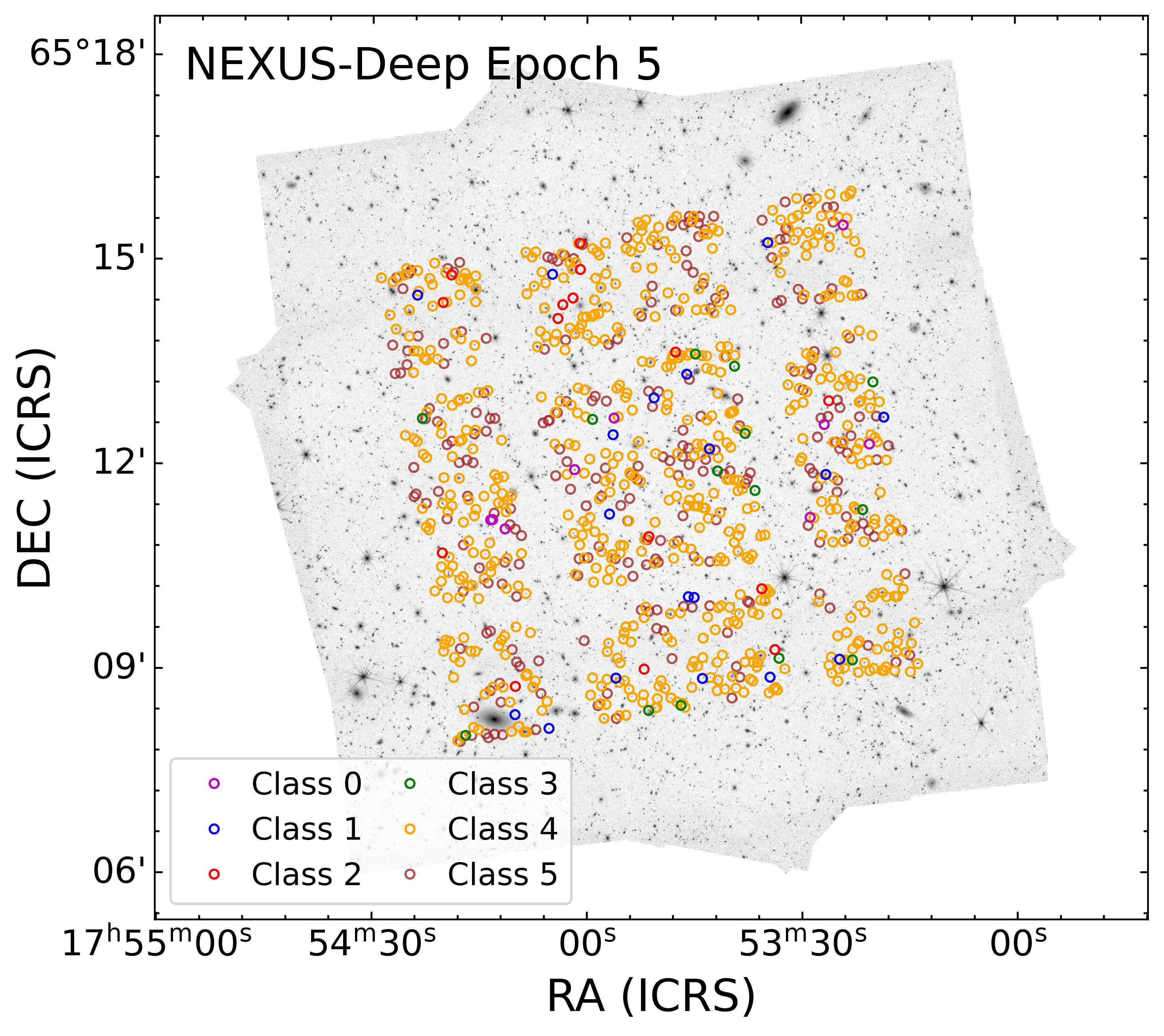}
    \includegraphics[width=0.46\linewidth]{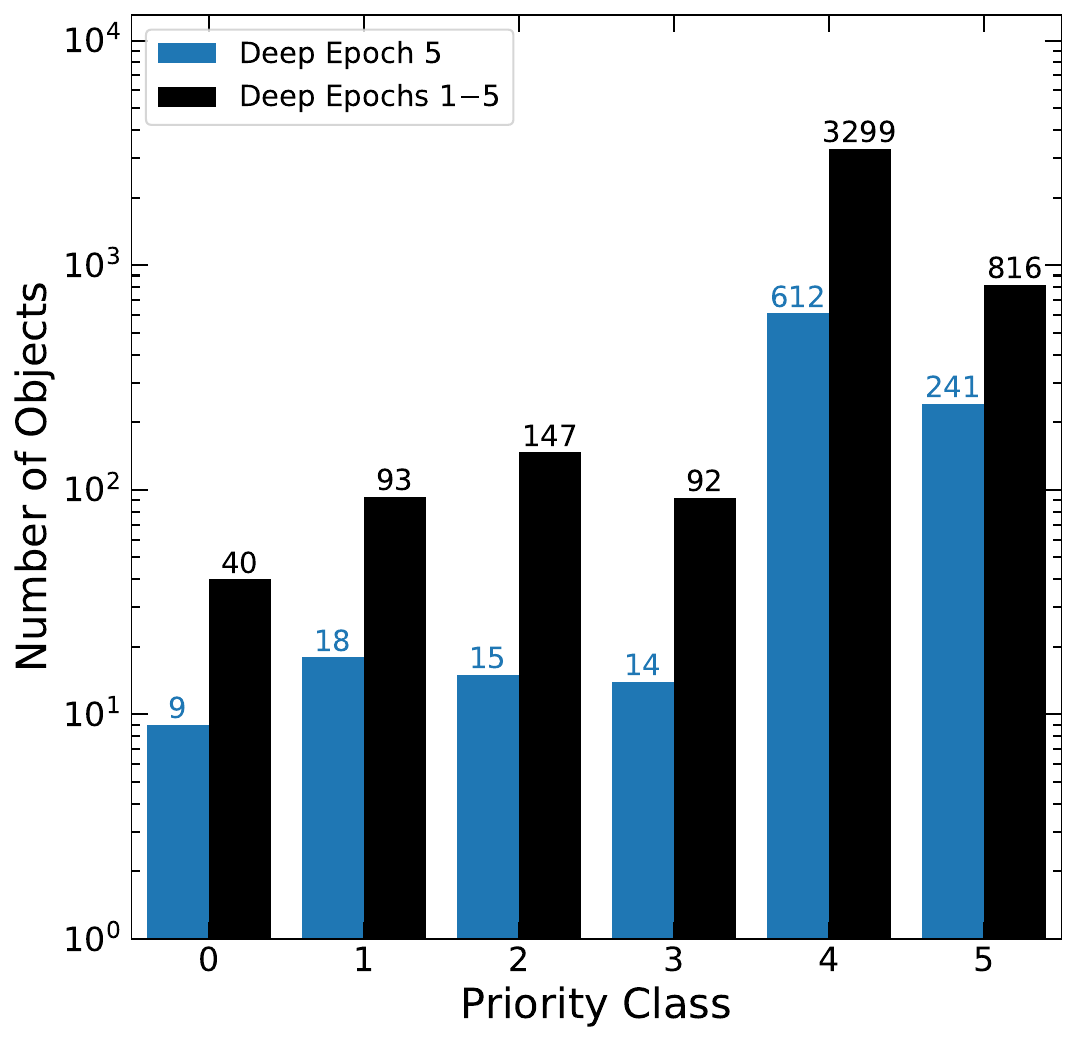}
    \includegraphics[width=0.46\linewidth]{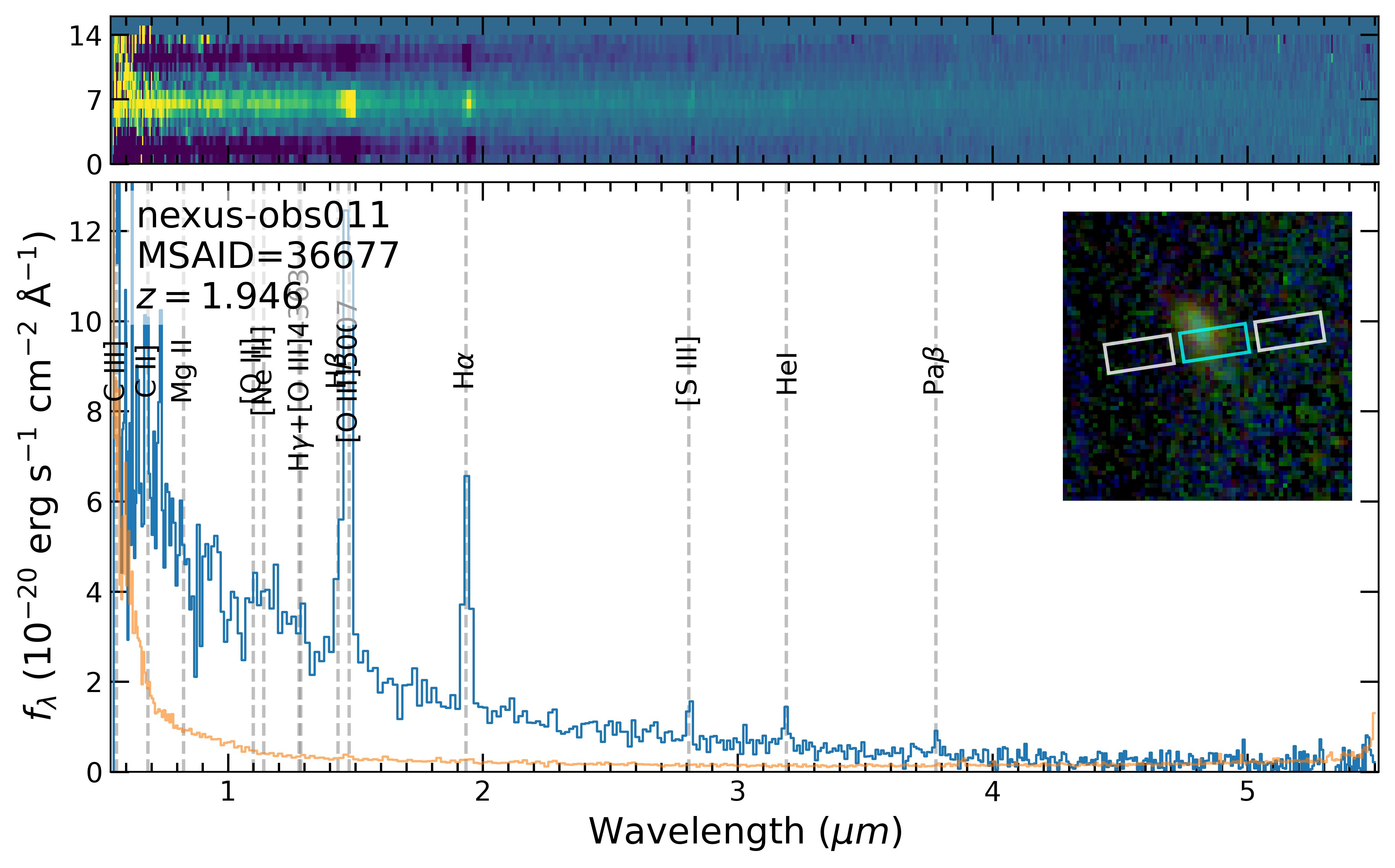}
    \includegraphics[width=0.46\linewidth]{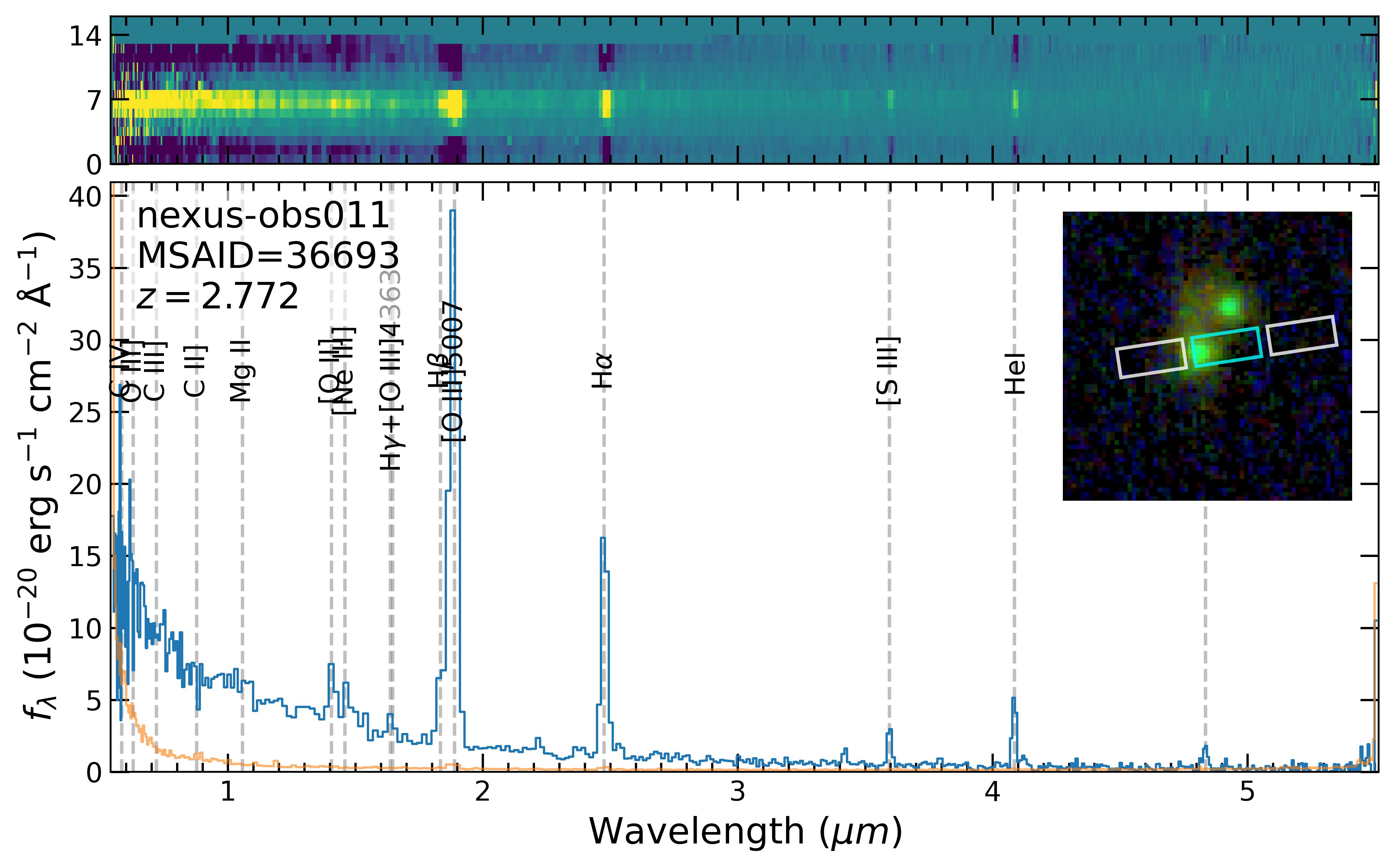}
    \caption{Same as Figure~\ref{fig:deep01} but for Deep Epoch 5.}
    \label{fig:deep05}
\end{figure*}

\subsection{Deep6: 2026/3/28}
{\textbf{Summary of changes}: To make use of additional available time due to evolution of APT versions, we increased the exposure time of Deep Epoch 6 NIRSpec MOS observations. Each pointing now uses 37 Groups and 2 Integrations (instead of 43 Groups and 1 Integration) with the same two-shutter slitlet nodding, which increases the effective exposure time per target from 21.4 to 37.0 min. The coordinated parallel NIRCam imaging was correspondingly changed from the MEDIUM8 readout pattern with 6 Groups and 1 Integration (1245.5\,s) to SHALLOW4 with 10 Groups and 2 Integrations (2125.9\,s), both with 2 dithers.}

The top panels in Figure~\ref{fig:deep06} show the targets layout of Deep Epoch 6 and the histogram of targets observed so far as a function of priority class. We observed 10 class=0 objects, 25 class=1 objects, 6 class=2 objects, 13 class=3 objects, 596 class=4 objects, 225 class=5 objects, with 875 objects in total. 52 contaminants also fell within open slitlets.

In Deep Epoch 6, we highlight a $z=3.462$ LRD with a prominent host galaxy (MSAID=18979) and a $z=0.759$ supernova (MSAID=36709). Source 18979 was identified as an LRD by \citet{nexus_lrd}, and its host galaxy is clearly resolved in the NIRCam color composite around the compact red nucleus. Its PRISM spectrum shows strong \OII, \NeIII, \Hb, \OIII, and \Ha\ emission lines. As a Class 1 target, source 18979 has been observed in multiple Deep epochs for spectroscopic monitoring. A detailed analysis of this object is presented in \citet{nexus_prism_lrd}. Source 36709 is the spectrum of the transient AT~2026rga, which was discovered in the Deep Epoch 5 NIRCam imaging and reported in \citet{nexus_transients}. The transient lies far from its host galaxy (``hostless'' by our $>1''$ criterion), so its spectrum is free of host contamination. Many emission lines show P-Cygni profiles, most notably the broad \Ha\ line, which classifies AT~2026rga as a Type IIP/IIL supernova. This object demonstrates the capability of NEXUS-Deep to discover transients in imaging and obtain their spectroscopic classification in the next epoch, made possible by the 2-month cadence.

\begin{figure*}[ht]
    \centering
    \includegraphics[width=0.5\linewidth]{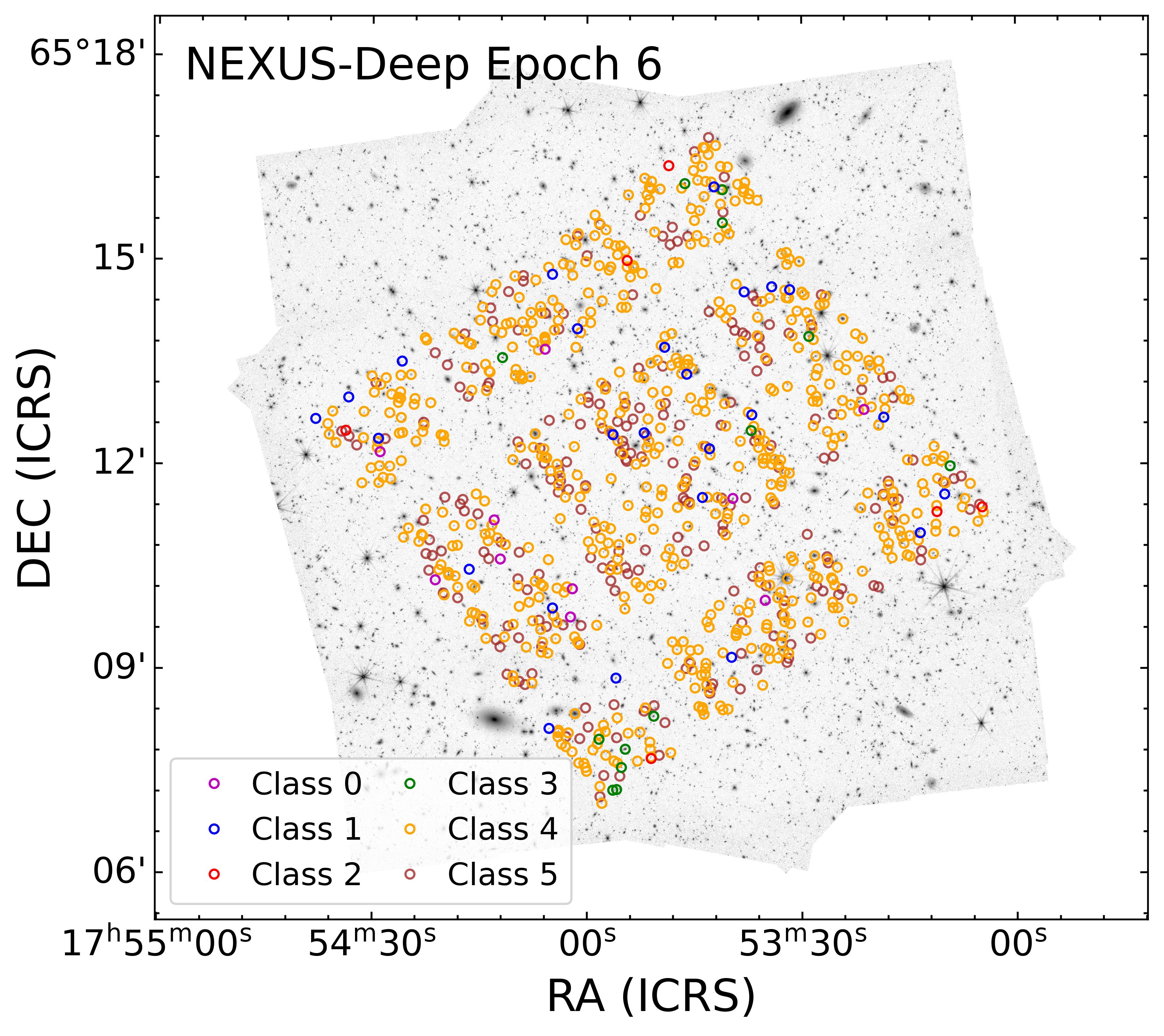}
    \includegraphics[width=0.46\linewidth]{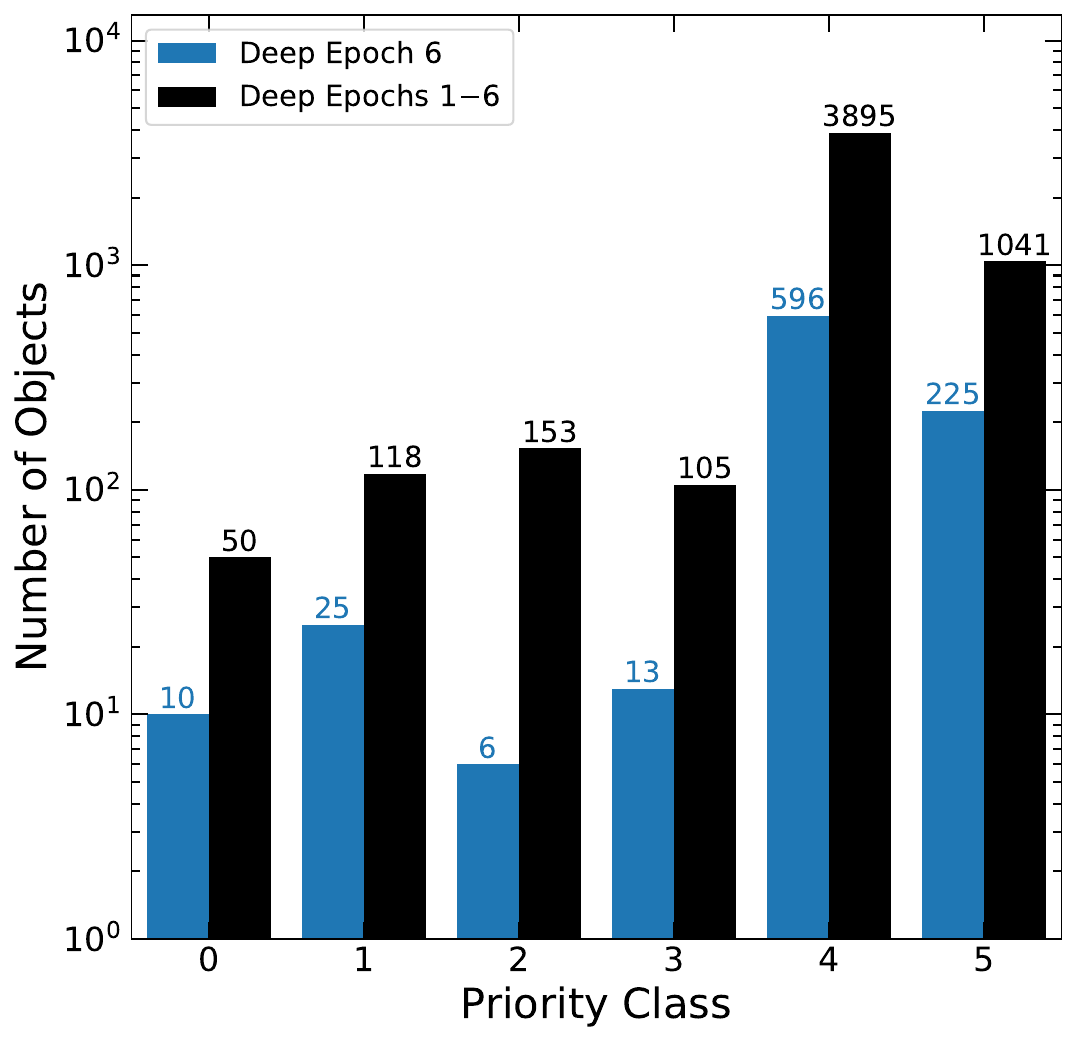}
    \includegraphics[width=0.46\linewidth]{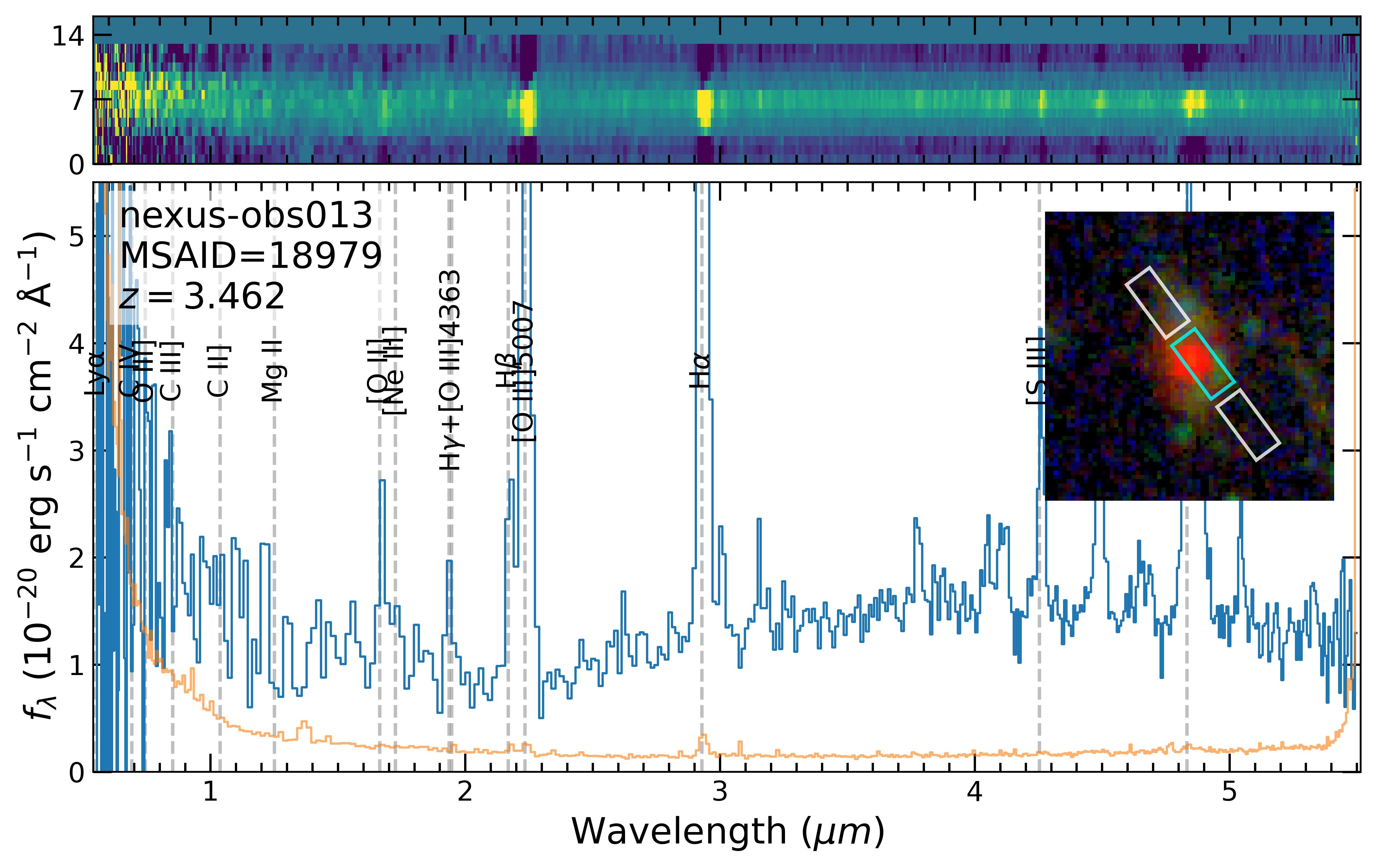}
    \includegraphics[width=0.46\linewidth]{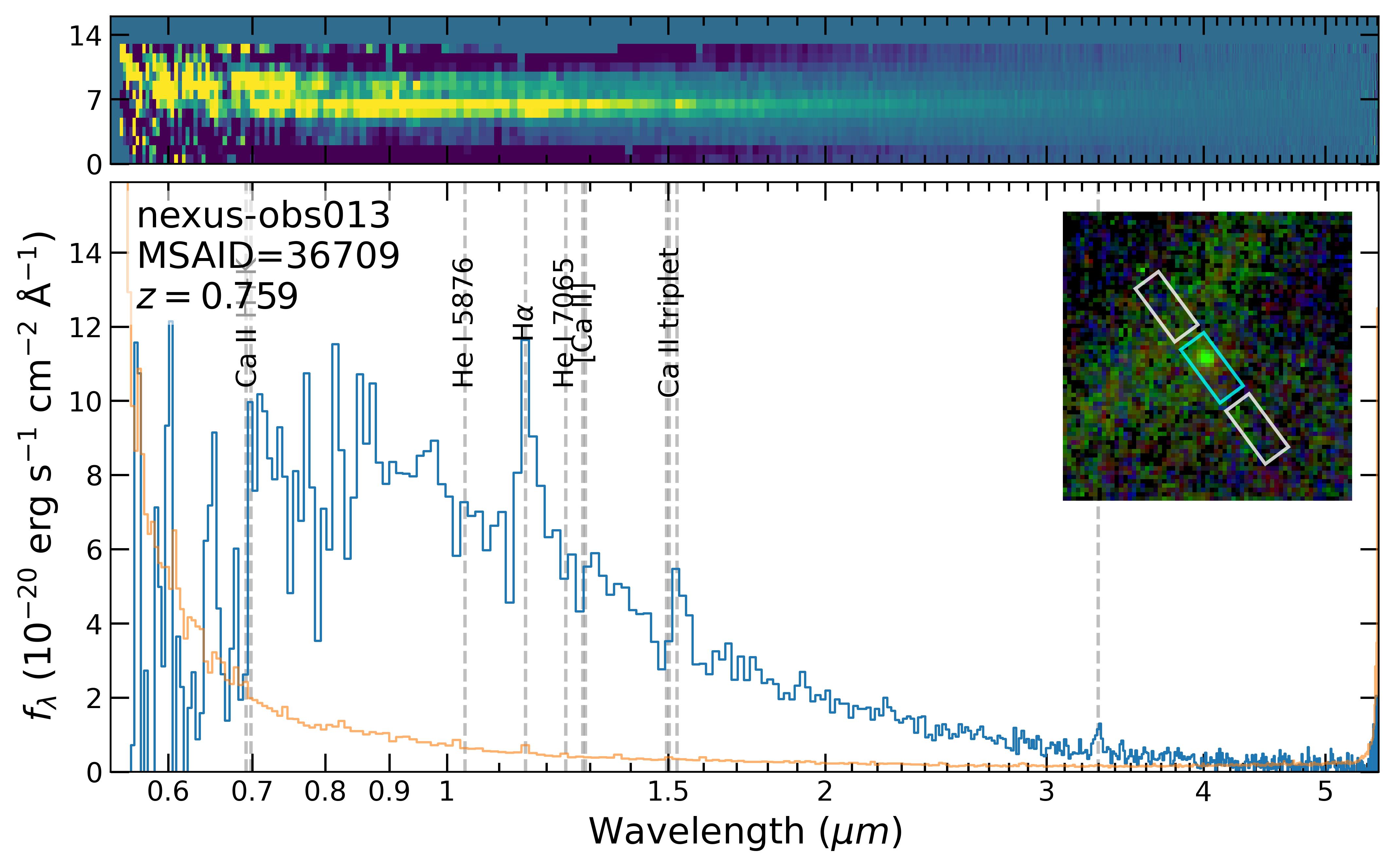}
    \caption{Same as Figure~\ref{fig:deep01} but for Deep Epoch 6.}
    \label{fig:deep06}
\end{figure*}

\subsection{Deep7: 2026/5/28}
{\textbf{Summary of changes}: 1. We added 23 bright (F444W$<$26 mag) photometric LRD candidates from an updated LRD candidate catalog (Pan~et~al in prep.) to Class 1. 2. We reorganized the transient target list. Duplicate transients discovered independently in different epochs were merged, and spurious detections were removed based on their morphology, their detection in individual exposures, and their presence in nearly contemporaneous medium-band imaging when available. The remaining transients were assigned new MSAIDs. For each transient, we aim to obtain two spectra: one while the transient is still visible, and a second after it has faded if the ``on''-state spectrum is contaminated by its host galaxy. 3. We introduced a new subclass, Class=0.5, for the underlying host galaxy at the location of a transient, with weight=10. 4. Starting from Cycle 4 (the second year of NEXUS-Deep), we systematically increased the weights of Class=0.1 transients from 10 to 20 for 26$<$F200W or F444W$<$27, and from 1 to 4 for 27$<$F200W or F444W$<$28, so that faint transients with strong emission lines (e.g., Type II supernovae) are more likely to receive a slit.}

The top panels in Figure~\ref{fig:deep07} show the targets layout of Deep Epoch 7 and the histogram of targets observed so far as a function of priority class. We observed 16 class=0 objects, 31 class=1 objects, 3 class=2 objects, 4 class=3 objects, 500 class=4 objects, 265 class=5 objects, with 819 objects in total. 40 contaminants also fell within open slitlets.

Starting from Deep Epoch 7, we no longer highlight individual objects in these notes. Spectra of all observed targets are available in the quick data release.

\begin{figure*}[ht]
    \centering
    \includegraphics[width=0.5\linewidth]{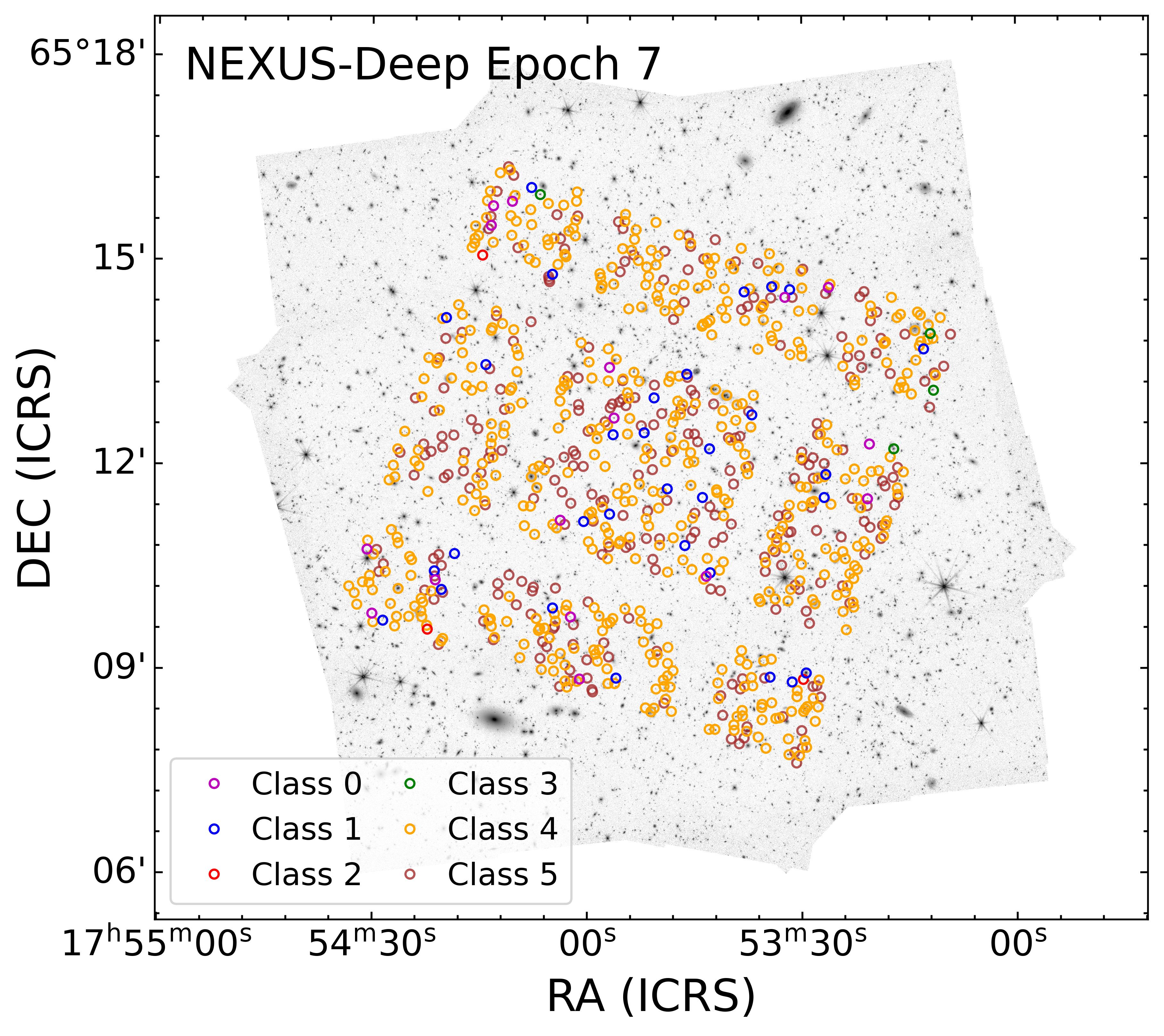}
    \includegraphics[width=0.46\linewidth]{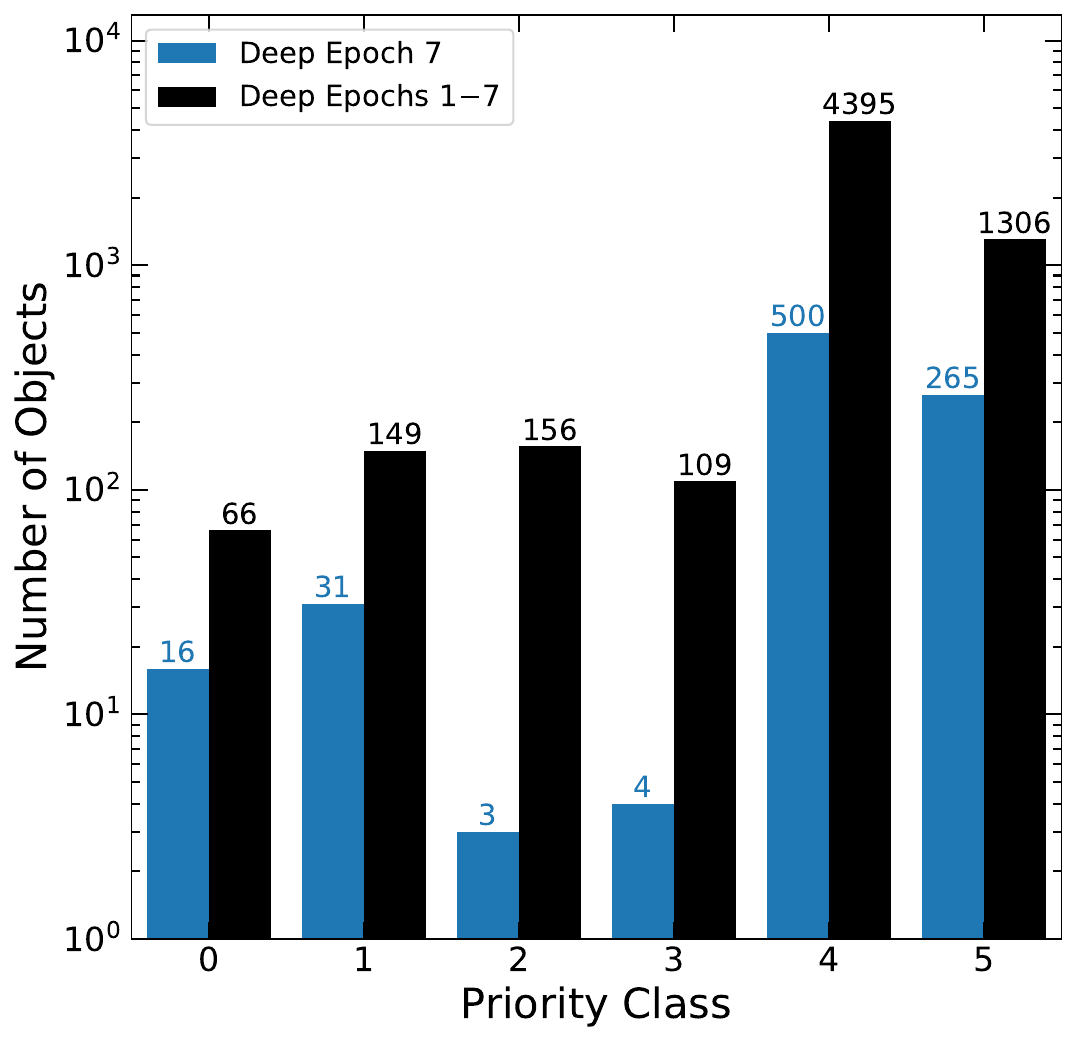}
    \caption{Same as Figure~\ref{fig:deep01} but for Deep Epoch 7.}
    \label{fig:deep07}
\end{figure*}

\subsection{Deep8: 2026/8/7}
{\textbf{Summary of changes}: 1. Class=0.5 now includes the assigned host galaxies of all transients that do not yet have a spectroscopic redshift (weight=10), rather than only the galaxy directly underlying the transient position. 2. Due to a NIRCam anomaly, the observing window of Deep Epoch 8 was moved from July 24--30, 2026 to August 3--14, 2026, with a corresponding change in the NIRSpec aperture PA.}

The top panels in Figure~\ref{fig:deep08} show the targets layout of Deep Epoch 8 and the histogram of targets observed so far as a function of priority class. We observed 21 class=0 objects, 27 class=1 objects, 5 class=2 objects, 7 class=3 objects, 515 class=4 objects, 257 class=5 objects, with 832 objects in total. 47 contaminants also fell within open slitlets.

\begin{figure*}[ht]
    \centering
    \includegraphics[width=0.5\linewidth]{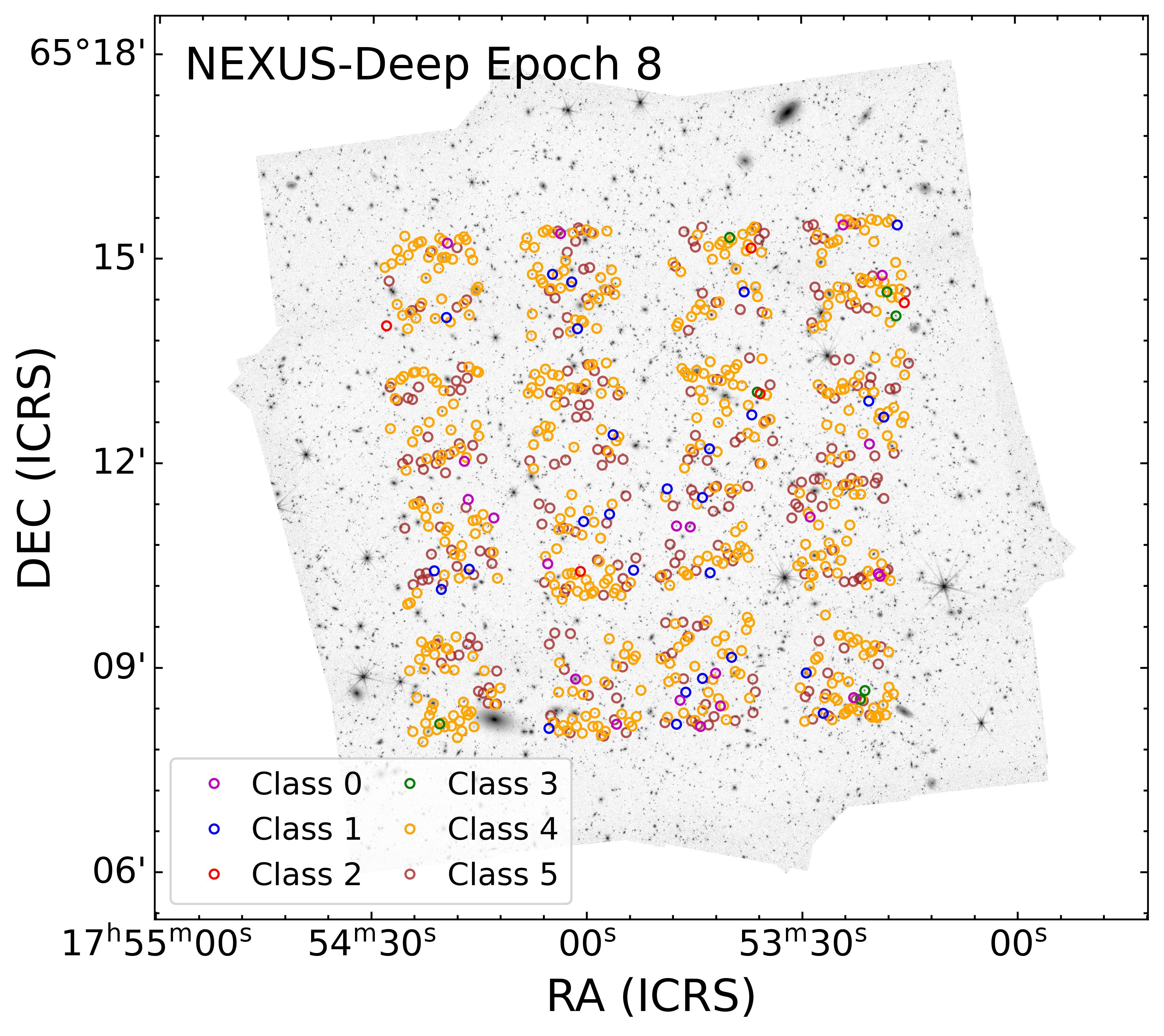}
    \includegraphics[width=0.46\linewidth]{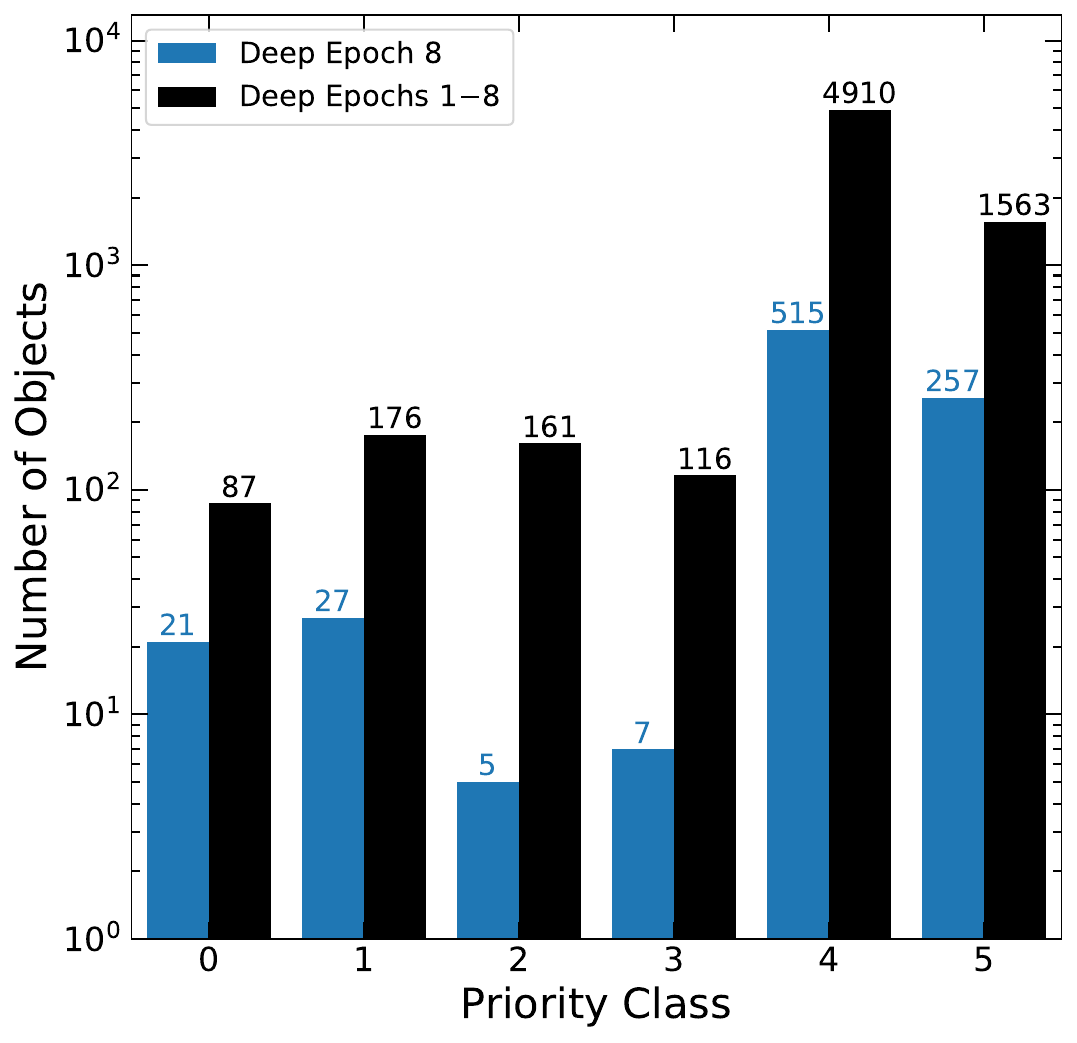}
    \caption{Same as Figure~\ref{fig:deep01} but for Deep Epoch 8.}
    \label{fig:deep08}
\end{figure*}

\acknowledgments

We thank the STScI JWST staff for the implementation of this multi-cycle program. Based on observations with the NASA/ESA/CSA James Webb Space Telescope obtained from the Barbara A. Mikulski Archive at the Space Telescope Science Institute, which is operated
by the Association of Universities for Research in Astronomy, Incorporated, under NASA contract NAS5-03127. Support for Program number JWST-GO-05105 was provided through a grant from the STScI under NASA contract NAS5-03127. 



\bibliographystyle{aasjournal}
\bibliography{refs.bib}

@ARTICLE{Greene+2024,
       author = {{Greene}, Jenny E. and {Labbe}, Ivo and {Goulding}, Andy D. and {Furtak}, Lukas J. and {Chemerynska}, Iryna and {Kokorev}, Vasily and {Dayal}, Pratika and {Volonteri}, Marta and {Williams}, Christina C. and {Wang}, Bingjie and {Setton}, David J. and {Burgasser}, Adam J. and {Bezanson}, Rachel and {Atek}, Hakim and {Brammer}, Gabriel and {Cutler}, Sam E. and {Feldmann}, Robert and {Fujimoto}, Seiji and {Glazebrook}, Karl and {de Graaff}, Anna and {Khullar}, Gourav and {Leja}, Joel and {Marchesini}, Danilo and {Maseda}, Michael V. and {Matthee}, Jorryt and {Miller}, Tim B. and {Naidu}, Rohan P. and {Nanayakkara}, Themiya and {Oesch}, Pascal A. and {Pan}, Richard and {Papovich}, Casey and {Price}, Sedona H. and {van Dokkum}, Pieter and {Weaver}, John R. and {Whitaker}, Katherine E. and {Zitrin}, Adi},
        title = "{UNCOVER Spectroscopy Confirms the Surprising Ubiquity of Active Galactic Nuclei in Red Sources at z > 5}",
      journal = {\apj},
         year = 2024,
        month = mar,
       volume = {964},
       number = {1},
          eid = {39},
        pages = {39},
          doi = {10.3847/1538-4357/ad1e5f},
archivePrefix = {arXiv},
       eprint = {2309.05714},
 primaryClass = {astro-ph.GA},
       adsurl = {https://ui.adsabs.harvard.edu/abs/2024ApJ...964...39G}
}

@ARTICLE{EAZY,
       author = {{Brammer}, Gabriel B. and {van Dokkum}, Pieter G. and {Coppi}, Paolo},
        title = "{EAZY: A Fast, Public Photometric Redshift Code}",
      journal = {\apj},
         year = 2008,
        month = oct,
       volume = {686},
       number = {2},
        pages = {1503-1513},
          doi = {10.1086/591786},
archivePrefix = {arXiv},
       eprint = {0807.1533},
 primaryClass = {astro-ph},
       adsurl = {https://ui.adsabs.harvard.edu/abs/2008ApJ...686.1503B}
}

@ARTICLE{2023Natur.619..716C,
       author = {{Carnall}, Adam C. and {McLure}, Ross J. and {Dunlop}, James S. and {McLeod}, Derek J. and {Wild}, Vivienne and {Cullen}, Fergus and {Magee}, Dan and {Begley}, Ryan and {Cimatti}, Andrea and {Donnan}, Callum T. and {Hamadouche}, Massissilia L. and {Jewell}, Sophie M. and {Walker}, Sam},
        title = "{A massive quiescent galaxy at redshift 4.658}",
      journal = {\nat},
         year = 2023,
        month = jul,
       volume = {619},
       number = {7971},
        pages = {716-719},
          doi = {10.1038/s41586-023-06158-6},
archivePrefix = {arXiv},
       eprint = {2301.11413},
 primaryClass = {astro-ph.GA},
       adsurl = {https://ui.adsabs.harvard.edu/abs/2023Natur.619..716C}
}

@ARTICLE{Taylor+2024,
       author = {{Taylor}, Anthony J. and {Finkelstein}, Steven L. and {Kocevski}, Dale D. and {Jeon}, Junehyoung and {Bromm}, Volker and {Amor{\'\i}n}, Ricardo O. and {Arrabal Haro}, Pablo and {Backhaus}, Bren E. and {Bagley}, Micaela B. and {Banados}, Eduardo and {Bhatawdekar}, Rachana and {Brooks}, Madisyn and {Calabr{\`o}}, Antonello and {Ch{\'a}vez Ortiz}, {\'O}scar A. and {Cheng}, Yingjie and {Cleri}, Nikko J. and {Cole}, Justin W. and {Davis}, Kelcey and {Dickinson}, Mark and {Donnan}, Callum and {Dunlop}, James S. and {Ellis}, Richard S. and {Fern{\'a}ndez}, Vital and {Fontana}, Adriano and {Fujimoto}, Seiji and {Giavalisco}, Mauro and {Grazian}, Andrea and {Guo}, Jingsong and {Hathi}, Nimish P. and {Holwerda}, Benne W. and {Hirschmann}, Michaela and {Inayoshi}, Kohei and {Kartaltepe}, Jeyhan S. and {Khusanova}, Yana and {Koekemoer}, Anton M. and {Kokorev}, Vasily and {Larson}, Rebecca L. and {Leung}, Gene C.~K. and {Lucas}, Ray A. and {McLeod}, Derek J. and {Napolitano}, Lorenzo and {Onoue}, Masafusa and {Pacucci}, Fabio and {Papovich}, Casey and {P{\'e}rez-Gonz{\'a}lez}, Pablo G. and {Pirzkal}, Nor and {Somerville}, Rachel S. and {Trump}, Jonathan R. and {Wilkins}, Stephen M. and {Yung}, L.~Y. Aaron and {Zhang}, Haowen},
        title = "{Broad-line AGNs at 3.5 < z < 6: The Black Hole Mass Function and a Connection with Little Red Dots}",
      journal = {\apj},
         year = 2025,
        month = jun,
       volume = {986},
       number = {2},
          eid = {165},
        pages = {165},
          doi = {10.3847/1538-4357/add15b},
archivePrefix = {arXiv},
       eprint = {2409.06772},
 primaryClass = {astro-ph.GA},
       adsurl = {https://ui.adsabs.harvard.edu/abs/2025ApJ...986..165T}
}

@ARTICLE{LeiHu2024,
       author = {{Hu}, Lei and {Wang}, Lifan},
        title = "{Differencing and Coadding JWST Images with Matched Point-spread Function}",
      journal = {\aj},
         year = 2024,
        month = may,
       volume = {167},
       number = {5},
          eid = {231},
        pages = {231},
          doi = {10.3847/1538-3881/ad36cb},
archivePrefix = {arXiv},
       eprint = {2309.09143},
 primaryClass = {astro-ph.IM},
       adsurl = {https://ui.adsabs.harvard.edu/abs/2024AJ....167..231H}
}

@software{msaexp,
       author = {{Brammer}, Gabriel},
        title = "{msaexp: NIRSpec analyis tools}",
         year = 2023,
        month = sep,
          eid = {10.5281/zenodo.7299500},
          doi = {10.5281/zenodo.7299500},
      version = {0.6.17},
    publisher = {Zenodo},
       adsurl = {https://ui.adsabs.harvard.edu/abs/2022zndo...7299500B}
}

@ARTICLE{QSO_selection,
       author = {{Wu}, Xue-Bing and {Hao}, Guoqiang and {Jia}, Zhendong and {Zhang}, Yanxia and {Peng}, Nanbo},
        title = "{SDSS Quasars in the WISE Preliminary Data Release and Quasar Candidate Selection with Optical/Infrared Colors}",
      journal = {\aj},
         year = 2012,
        month = aug,
       volume = {144},
       number = {2},
          eid = {49},
        pages = {49},
          doi = {10.1088/0004-6256/144/2/49},
archivePrefix = {arXiv},
       eprint = {1204.6197},
 primaryClass = {astro-ph.CO},
       adsurl = {https://ui.adsabs.harvard.edu/abs/2012AJ....144...49W}
}

@ARTICLE{desi_dr1,
       author = {{DESI Collaboration} and {Abdul-Karim}, M. and {Adame}, A.~G. and {Aguado}, D. and {Aguilar}, J. and {Ahlen}, S. and {Alam}, S. and {Aldering}, G. and {Alexander}, D.~M. and {Alfarsy}, R. and {Allen}, L. and {Allende Prieto}, C. and {Alves}, O. and {Anand}, A. and {Andrade}, U. and {Armengaud}, E. and {Avila}, S. and {Aviles}, A. and {Awan}, H. and {Bailey}, S. and {Baleato Lizancos}, A. and {Ballester}, O. and {Bault}, A. and {Bautista}, J. and {BenZvi}, S. and {Beraldo e Silva}, L. and {Bermejo-Climent}, J.~R. and {Beutler}, F. and {Bianchi}, D. and {Blake}, C. and {Blum}, R. and {Bolton}, A.~S. and {Bonici}, M. and {Brieden}, S. and {Brodzeller}, A. and {Brooks}, D. and {Buckley-Geer}, E. and {Burtin}, E. and {Canning}, R. and {Carnero Rosell}, A. and {Carr}, A. and {Carrilho}, P. and {Casas}, L. and {Castander}, F.~J. and {Cereskaite}, R. and {Cervantes-Cota}, J.~L. and {Chaussidon}, E. and {Chaves-Montero}, J. and {Chen}, S. and {Chen}, X. and {Claybaugh}, T. and {Cole}, S. and {Cooper}, A.~P. and {Cousinou}, M.-C. and {Cuceu}, A. and {Davis}, T.~M. and {Dawson}, K.~S. and {de Belsunce}, R. and {de la Cruz}, R. and {de la Macorra}, A. and {de Mattia}, A. and {Deiosso}, N. and {Della Costa}, J. and {Demina}, R. and {Demirbozan}, U. and {DeRose}, J. and {Dey}, A. and {Dey}, B. and {Ding}, J. and {Ding}, Z. and {Doel}, P. and {Douglass}, K. and {Dowicz}, M. and {Ebina}, H. and {Edelstein}, J. and {Eisenstein}, D.~J. and {Elbers}, W. and {Emas}, N. and {Escoffier}, S. and {Fagrelius}, P. and {Fan}, X. and {Fanning}, K. and {Fawcett}, V.~A. and {Fern{\'a}ndez-Garc{\'i}a}, E. and {Ferraro}, S. and {Findlay}, N. and {Font-Ribera}, A. and {Forero-Romero}, J.~E. and {Forero-S{\'a}nchez}, D. and {Frenk}, C.~S. and {G{\"a}nsicke}, B.~T. and {Galbany}, L. and {Garc{\'i}a-Bellido}, J. and {Garcia-Quintero}, C. and {Garrison}, L.~H. and {Gazta{\~n}aga}, E. and {Gil-Mar{\'i}n}, H. and {Gnedin}, O.~Y. and {Gontcho}, S. Gontcho A and {Gonzalez-Morales}, A.~X. and {Gonzalez-Perez}, V. and {Gordon}, C. and {Graur}, O. and {Green}, D. and {Gruen}, D. and {Gsponer}, R. and {Guandalin}, C. and {Gutierrez}, G. and {Guy}, J. and {Hahn}, C. and {Han}, J.~J. and {Han}, J. and {He}, S. and {Herrera-Alcantar}, H.~K. and {Honscheid}, K. and {Hou}, J. and {Howlett}, C. and {Huterer}, D. and {Ir{\v{s}}i{\v{c}}}, V. and {Ishak}, M. and {Jacques}, A. and {Jimenez}, J. and {Jing}, Y.~P. and {Joachimi}, B. and {Joudaki}, S. and {Joyce}, R. and {Jullo}, E. and {Juneau}, S. and {Kara{\c{c}}ayl{\i}}, N.~G. and {Karim}, T. and {Kehoe}, R. and {Kent}, S. and {Khederlarian}, A. and {Kirkby}, D. and {Kisner}, T. and {Kitaura}, F.-S. and {Kizhuprakkat}, N. and {Kong}, H. and {Koposov}, S.~E. and {Kremin}, A. and {Krolewski}, A. and {Lahav}, O. and {Lai}, Y. and {Lamman}, C. and {Lan}, T.-W. and {Landriau}, M. and {Lang}, D. and {Lange}, J.~U. and {Lasker}, J. and {Le Goff}, J.~M. and {Le Guillou}, L. and {Leauthaud}, A. and {Levi}, M.~E. and {Li}, S. and {Li}, T.~S. and {Lodha}, K. and {Lokken}, M. and {Luo}, Y. and {Magneville}, C. and {Manera}, M. and {Manser}, C.~J. and {Margala}, D. and {Martini}, P. and {Maus}, M. and {McCullough}, J. and {McDonald}, P. and {Medina}, G.~E. and {Medina-Varela}, L. and {Meisner}, A. and {Mena-Fern{\'a}ndez}, J. and {Menegas}, A. and {Mezcua}, M. and {Miquel}, R. and {Montero-Camacho}, P. and {Moon}, J. and {Moustakas}, J. and {Mu{\~n}oz-Guti{\'e}rrez}, A. and {Mu{\~n}oz-Santos}, D. and {Myers}, A.~D. and {Myles}, J. and {Nadathur}, S. and {Najita}, J. and {Napolitano}, L. and {Newman}, J.~A. and {Nikakhtar}, F. and {Nikutta}, R. and {Niz}, G. and {Noriega}, H.~E. and {Padmanabhan}, N. and {Paillas}, E. and {Palanque-Delabrouille}, N. and {Palmese}, A. and {Pan}, J. and {Pan}, Z. and {Parkinson}, D. and {Peacock}, J. and {Percival}, W.~J. and {P{\'e}rez-Fern{\'a}ndez}, A. and {P{\'e}rez-R{\`a}fols}, I. and {Peterson}, P.},
        title = "{Data Release 1 of the Dark Energy Spectroscopic Instrument}",
      journal = {arXiv e-prints},
         year = 2025,
        month = mar,
          eid = {arXiv:2503.14745},
        pages = {arXiv:2503.14745},
          doi = {10.48550/arXiv.2503.14745},
archivePrefix = {arXiv},
       eprint = {2503.14745},
 primaryClass = {astro-ph.CO},
       adsurl = {https://ui.adsabs.harvard.edu/abs/2025arXiv250314745D}
}

@ARTICLE{nexus_lrd,
       author = {{Zhuang}, Ming-Yang and {Li}, Junyao and {Shen}, Yue and {Lin}, Xiaojing and {Shapley}, Alice E. and {Wang}, Feige and {Wu}, Qiaoya and {Yang}, Qian},
        title = "{NEXUS: A Spectroscopic Census of Broad-line AGNs and Little Red Dots at 3 {\ensuremath{\lesssim}} z {\ensuremath{\lesssim}} 6}",
      journal = {\apj},
         year = 2026,
        month = mar,
       volume = {999},
       number = {1},
          eid = {31},
        pages = {31},
          doi = {10.3847/1538-4357/ae3612},
archivePrefix = {arXiv},
       eprint = {2505.20393},
 primaryClass = {astro-ph.GA},
       adsurl = {https://ui.adsabs.harvard.edu/abs/2026ApJ...999...31Z}
}

@ARTICLE{nexus,
       author = {{Shen}, Yue and {Zhuang}, Ming-Yang and {Li}, Junyao and {Burgasser}, Adam J. and {Fan}, Xiaohui and {Greene}, Jenny E. and {Narayan}, Gautham and {Shapley}, Alice E. and {Sun}, Fengwu and {Wang}, Feige and {Yang}, Qian},
        title = "{NEXUS: the North ecliptic pole EXtragalactic Unified Survey}",
      journal = {arXiv e-prints},
         year = 2024,
        month = aug,
          eid = {arXiv:2408.12713},
        pages = {arXiv:2408.12713},
          doi = {10.48550/arXiv.2408.12713},
archivePrefix = {arXiv},
       eprint = {2408.12713},
 primaryClass = {astro-ph.GA},
       adsurl = {https://ui.adsabs.harvard.edu/abs/2024arXiv240812713S}
}

@ARTICLE{nexus_prism_lrd,
       author = {{Pan}, Zhiwei and {Zhuang}, Ming-Yang and {Shen}, Yue and {Wang}, Feige and {Greene}, Jenny E. and {Burgasser}, Adam J. and {Li}, Junyao and {Stone}, Zachary and {Venkatraman}, Padmavathi},
        title = "{NEXUS: Abundance, Environments, and Spectral Diversity of Little Red Dots from the NIRSpec MSA Sample}",
      journal = {arXiv e-prints},
         year = 2026,
        month = jun,
          eid = {arXiv:2606.09721},
        pages = {arXiv:2606.09721},
          doi = {10.48550/arXiv.2606.09721},
archivePrefix = {arXiv},
       eprint = {2606.09721},
 primaryClass = {astro-ph.GA},
       adsurl = {https://ui.adsabs.harvard.edu/abs/2026arXiv260609721P}
}

@ARTICLE{nexus_transients,
       author = {{Zhuang}, Ming-Yang and {Hu}, Lei and {Pierel}, Justin D.~R. and {Pan}, Zhiwei and {Shen}, Yue and {Burgasser}, Adam J. and {Fan}, Xiaohui and {Ho}, Luis C. and {Li}, Junyao and {Shapley}, Alice E. and {Stone}, Zachary and {Wang}, Feige and {Wang}, Lifan},
        title = "{NEXUS: Transient Searches and First Results from Year One Observations}",
      journal = {arXiv e-prints},
         year = 2026,
        month = sep,
          eid = {arXiv:2609.06985},
        pages = {arXiv:2609.06985},
          doi = {10.48550/arXiv.2609.06985},
archivePrefix = {arXiv},
       eprint = {2609.06985},
 primaryClass = {astro-ph.HE},
       adsurl = {https://ui.adsabs.harvard.edu/abs/2026arXiv260906985Z}
}

@ARTICLE{nexus_edr,
       author = {{Zhuang}, Ming-Yang and {Wang}, Feige and {Sun}, Fengwu and {Shen}, Yue and {Li}, Junyao and {Burgasser}, Adam J. and {Fan}, Xiaohui and {Greene}, Jenny E. and {Narayan}, Gautham and {Shapley}, Alice E. and {Yang}, Qian},
        title = "{NEXUS Early Data Release: NIRCam Imaging and WFSS Spectroscopy from the First (Partial) Wide Epoch}",
      journal = {\apjs},
         year = 2026,
        month = feb,
       volume = {282},
       number = {2},
          eid = {54},
        pages = {54},
          doi = {10.3847/1538-4365/ae2d05},
archivePrefix = {arXiv},
       eprint = {2411.06372},
 primaryClass = {astro-ph.GA},
       adsurl = {https://ui.adsabs.harvard.edu/abs/2026ApJS..282...54Z}
}

\end{document}